\documentclass[aps,10pt,pra,floatfix,twocolumn,superscriptaddress]{revtex4-1}
\usepackage{color}
\usepackage{ulem}
\usepackage{graphicx}
\usepackage{subfigure,amsmath, amsthm, amssymb}
\usepackage[colorlinks,citecolor=red]{hyperref}
\usepackage{braket}
\usepackage{enumitem}
\usepackage{natbib}

\newcommand{\be}{\begin{equation}}
\newcommand{\ee}{\end{equation}}
\newcommand{\bea}{\begin{eqnarray}}
\newcommand{\eea}{\end{eqnarray}}

\begin{document}

\title{Universal detection of entanglement in two-qubit states using only two copies}

\author{Suchetana Goswami}
\email{suchetana.goswami@gmail.com}
\affiliation{S. N. Bose National Centre for Basic Sciences, Salt Lake, Kolkata 700106,  India}

\author{Sagnik Chakraborty}
\email{csagnik@imsc.res.in}
\affiliation{Optics and Quantum Information Group, The Institute of Mathematical Sciences, C. I. T. Campus, Taramani, Chennai 600113, India}
\affiliation{Homi Bhabha National Institute, Training School Complex, Anushakti Nagar, Mumbai 400094, India}

\author{Sibasish Ghosh}
\email{sibasish@imsc.res.in}
\affiliation{Optics and Quantum Information Group, The Institute of Mathematical Sciences, C. I. T. Campus, Taramani, Chennai 600113, India}
\affiliation{Homi Bhabha National Institute, Training School Complex, Anushakti Nagar, Mumbai 400094, India}

\author{A. S. Majumdar}
\email{archan@bose.res.in}
\affiliation{S. N. Bose National Centre for Basic Sciences, Salt Lake, Kolkata 700106,  India}

\date{\today}
\begin{abstract}
We revisit the problem of detection of entanglement of an unknown two-qubit state using minimal resources.  Using weak values and just two copies of an arbitrary two-qubit state, we present a protocol where a post selection measurement in the computational basis provides enough information to identify if the state is entangled or not. Our protocol enables complete state  identification with a single-setting post selection measurement on two copies of the state. It follows that by restricting to pure states, the global interaction required for determining the weak values can be realized by local operations. We further show that our protocol is robust against errors arising from inappropriate global interactions applied during weak value determination.
\end{abstract}

\maketitle

\section{Introduction}

Ever since the coinage of the word ``entanglement" by Schr\"{o}dinger in 1935 closely following
the work of Einstein, Podolsky and Rosen (EPR) \cite{EPR_35}, discussion and debate about its nature and manifestation has continued to remain one of the most engaging issues in modern physics.
The paradox posed by EPR demonstrated for the first time the possibility of creating non-classical and nonlocal correlations with the help of entanglement, which Schr\"{o}dinger tried to explain in terms of quantum "steering" \cite{S_35}. Subsequently, the pioneering work of Bell \cite{B_64} paved the way for mathematically distinguishing quantum correlations  from those arising through a local realist description of physical phenomena. More recently, it has been realized that quantum correlations could be classified into hierarchical categories \cite{WJD_07, JWD_07} with entanglement being the weakest, followed by steering and Bell-nonlocality.

In present times, entanglement is regarded as the primary building block of quantum correlations,
leading to landmark discoveries in quantum information science \cite{enatnglement_review}. Numerous protocols have already been suggested, which use these correlations as resource and result in improvements, which no classical resource could achieve \cite{teleportation,dense,1sqkd,randomness, pironio2010random}. 
It has been realized \cite{WJD_07, JWD_07} that the nonlocal quantum correlations
responsible for steering and Bell-violation cannot exist without the presence of entanglement.
As a result of this, identification and quantification of quantum correlations, have become a topic of cutting edge research in various inter-disciplinary areas of physics, mathematics and computer science, as well.

In  quantum information theory, the way of identifying entanglement in a given bipartite state is through the {\it separability criterion} \cite{peres_sep,HORODECKI19961}. Though this
criterion is also helpful in quantification of entanglement \cite{negativity,log_negativity}, it is measurable only when full knowledge of the state is available. Such knowledge requires state tomography \cite{state_estimation_book}, which  is expensive in terms of resources required. On the other hand, there 
are methods based on direct measurement of observables (which are single setting measurements) such as {\it entanglement witnessing} \cite{T_00,LKCH_00,bruss2002} which have been experimentally realized \cite{experimental_witness,guhne2003experimental}. In addition, other schemes have been recently
proposed, such as {\it self-testing} protocols, which can identify individual entangled states giving rise to particular correlations in a given scenario \cite{MYS12,YN13,SH16,GBDSJM_18}. However,
all such methods suffer from the drawback of non-universality. For instance, for every entanglement witness (EW) there exists a class of entangled states, which it cannot detect \cite{guhne2009entanglement,LKCH_00}. This  prevents the use of any single EW to detect all entangled states. It is pertinent to note here that arranging a higher number of measurement settings is an expensive resource in experiments.

Entanglement detection in two-qubit  states has drawn renewed attention, as can be seen from 
several recent works \cite{recenttwoqubit3,*recenttwoqubit1,*recenttwoqubit2,*adhikari18}.
Our motivation for the present study is to reduce the resources required for identifying entanglement, and here 
we concern ourselves with the task of identification of entanglement in an unknown state.  
 In this context, Yu et al. \cite{ent_wit_rank2}  constructed an observable acting on four copies of any two-qubit state, 
that could detect entanglement for certain classes of two-qubit states. Augusiak et al. \cite{ADH_08} proposed the construction of an observable which acts on four copies of a two-qubit state and results in detection of all entangled states. Therefore, universal detection of entanglement could be done through measurement in a single setting, but the cost  is to supply multiple copies of the state.  
Further work in this direction \cite{girolami_2012,twocopy_concurrence_mixed3,dawei,heinosaari_2012},  has been performed to reduce the resources required for universal identification of entangled states. Girolami et al \cite{girolami_2012} proposed a method for identifying quantum correlations in two-qubit states through measurement of seven observables on four copies, where the observables are local in the Alice-Bob cut (the two parties sharing the bipartite state). It has been shown \cite{dawei, heinosaari_2012} that any universal entanglement detection scheme on a single copy of a state, has to be necessarily a state tomography process. Recently, the protocol in \cite{ADH_08} was extended to the completely device independent scenario \cite{MG_17}.

In the present work we propose a protocol where universal detection of entanglement is possible in a single  measurement setting on just {\it two} copies of any two-qubit state, using weak values.
The idea of weak measurement was first proposed by Aharonov et al. in \cite{Vaidman_weak}, to show that an experimental outcome outside the eigenvalue spectrum of an observable could be obtained if a sufficiently weak coupling of the system and the apparatus along with post-selection is employed. Weak measurements have  been utilized in several interesting applications such as observations of spin Hall effect \cite{HK_08}, trajectories of photons \cite{KBRSM_11}, direct measurement of the quantum wave function \cite{LSPSB_11}, and measurement of ultrasmall time delays of light \cite{SB_13}. The technique of weak measurement and reversal has also been used in the preservation of entanglement \cite{KU_99, KJ_06, KLKK_12, MXA_12}, teleportation fidelity \cite{PM_13} and steerability \cite{DGPM_17} through noisy channels. Detection of weak value  has  been found to be useful in observing geometric phase \cite{w_value_geometric}, non-Hermitian operators \cite{w_value-non-Herm} and
  quantum state \cite{w_value_state1,w_value_state2,w_value_state3}.

Here we show that our protocol of entanglement detection using weak values on two copies of an
arbitrary two-qubit state results in complete identification of the  
state, i.e., state tomography, in the similar fashion as in \cite{dawei,heinosaari_2012}. 
Note that, a number of attempts \cite{twocopy_concurrence_pure,twocopy_concurrence_mixed1,twocopy_concurrence_mixed2,twocopy_concurrence_mixed3} were made to measure concurrence \cite{coffman,wootters_concurrence} of two qubit states through measurement of a single observable on two copies of the state. Although, for pure states \cite{twocopy_concurrence_pure} such observables could be found, only estimates could be given for mixed states \cite{twocopy_concurrence_mixed1,twocopy_concurrence_mixed2,twocopy_concurrence_mixed3}. In this regard, our result provides a solution to this problem, as complete identification of two-qubit states, obtained through our protocol, also imply measurement of concurrence for any two-qubit state  using two copies. 
We further show that on restricting the set of 
states to just pure states, the weak interaction necessary in our protocol, can be realized through local operations on each of the qubits. Finally, we also show that our protocol is robust to errors arising from inappropriate choice of weak interaction between two copies of the two-qubit states.
 
 The plan of this paper is as follows.
In Sec \ref{Background} we discuss the preliminaries required for the analysis, before presenting our protocol in Sec \ref{method}. In Sec \ref{local impementation} we discuss possible implementation of our scheme through local operations. In Sec \ref{robust} we demonstrate the robustness of our protocol, before concluding in Sec \ref{conclusion}.

\section{Background}
\label{Background}
For any Hilbert spaces $\mathcal{H}$, let the space of all linear operators be denoted by $\mathcal{L}(\mathcal{H})$, and the set of all density matrices be $\mathcal{P}_+(\mathcal{H})$. Now, consider two parties, Alice and Bob, each separately possessing a two-level quantum system (qubit) with Hilbert spaces $\mathcal{H}_A$  and $\mathcal{H}_B$, respectively. Also consider the pointer system of the measuring apparatus to be a quantum system with Hilbert space $\mathcal{H}_{E}$.
Now, any bipartite quantum state,
that can be written as a convex mixture of product states is called a {\it separable} state i.e.
\begin{eqnarray}
\rho_{sep} = \sum_{i} p_{i} \rho_{i}^{A} \otimes \rho_{i}^{B}
\label{ent_state}
\end{eqnarray}
for any $\rho_{i}^{A}\in\mathcal{P}_+(\mathcal{H}_A)$ , $\rho_{i}^{B}\in\mathcal{P}_+(\mathcal{H}_B)$ and  probability distribution $\{p_i\}_i$. Any state $\rho\in\mathcal{P}_+(\mathcal{H}_A\otimes\mathcal{H}_B)$ which is not separable is called an {\it entangled} state. Note that, as $\rho\in\mathcal{L}(\mathcal{H}_A\otimes\mathcal{H}_B)$, it can always be decomposed as,
\begin{eqnarray}
\rho=\sum_{ijkl} p_{kl}^{ij} \ket{i} \bra{j} \otimes \ket{k} \bra{l}
\label{ppt1}
\end{eqnarray}
where $\{\ket{i}\}_i$ denotes an orthonormal basis in each of the subsystem Hilbert spaces $\mathcal{H}_A$ and $\mathcal{H}_B$. Using this decomposition, we can define the partial transpose of $\rho$, with respect to the  subsystem $B$, in the following way,
\begin{eqnarray}
\rho^{T_{B}}=\sum_{ijkl} p_{lk}^{ij} \ket{i} \bra{j} \otimes \ket{k} \bra{l}
\label{ppt2}
\end{eqnarray}
Note that, we can similarly define $\rho^{T_A}$ and $\rho^{T_{B}}=({\rho^{T_{A}}})^T$, where $\bullet^T$ denotes transposition. Now, we can present the separability criteria, as mentioned in the previous section. Any two qubit state $\rho\in\mathcal{P}_+(\mathcal{H}_A\otimes\mathcal{H}_B)$ is separable \cite{ADH_08,sanpera_two_qubit,Verstraete_two_qubit} if and only if,
\begin{eqnarray}
det~ \big(\rho^{T_{B}}\big) \geq 0 
\label{det_sep}
\end{eqnarray} 
where $det(A)$ represents determinant of a matrix $A$. This criterion can also be linked to the quantification of entanglement in terms of concurrence \cite{wootters_concurrence}.

Now we briefly illustrate the idea of weak measurement and weak values. In the theory of weak measurements \cite{Vaidman_weak,weak_review}, the pointer system of the measuring apparatus is kept in an initial state $\phi_{in}\in\mathcal{P}_+(\mathcal{H}_E)$ and a quantum system is {\it pre-selected} in a state $\rho\in\mathcal{P}_+(\mathcal{H})$. Then, the joint system-pointer state is evolved through a {\it weak interaction} generated by a Hamiltonian $\epsilon H\otimes P_x$, where  $H$ is the Hamiltonian associated with the system, $P_x$ is the momentum operator of the pointer system, and $\epsilon$ is a small positive number representing the {\it weakness} of interaction. Following this, a strong {\it post-selective} measurement is performed on the weakly evolved state of the system in a basis $\{\ket{u_k}\}_k$, where $\ket{u_k}\in\mathcal{H}$, which results in the pointer state $\phi^k_f\in\mathcal{P}_+(\mathcal{H}_E)$ for each $k$, where
\begin{eqnarray}
\label{weakpointer}
 \phi^k_f\approx \bra{u_i}\rho\ket{u_i}e^{-i\epsilon \langle H \rangle^{(k)}_{\rho}~P_x} \phi_{in}~ e^{i\epsilon  \langle H \rangle^{(k)}_{\rho}~P_x}
\end{eqnarray}
where $\langle H \rangle^{(k)}_{\rho}$ are the weak values, given by,
\begin{eqnarray}
\langle H \rangle^{(k)}_{\rho}= \frac{tr\big[H \rho \ket{u_k} \bra{u_k}\big]}{tr\big[\rho \ket{u_k} \bra{u_k}\big]}.
\label{weakvalue}
\end{eqnarray}
Note that Eq. (\ref{weakpointer}) can be derived only under the approximation that $\epsilon$ is very small. 
For measuring $\langle H \rangle^{(k)}_{\rho}$ certain properties of the position and momentum wave function of $\phi^k_f$ needs to be observed. As mentioned in Ref \cite{Jozsacomplex}, these properties include 
 shift in expectation value of the position and momentum wavefunction  compared to their initial values, variance of the momentum wave function, rate of change of the position wavefunction, and strength of the weak interaction i e., $\epsilon$.
 A detailed analysis on this technique is provided in section II of  \cite{Jozsacomplex}. 
Also recently, real and imaginary parts of a weak value was detected by using Laguerre-Gaussian modes \cite{w_value_detection_single} in the pointer state. For a detailed discussion on weak values, refer to \cite{weak_review}.

\section{Entanglement witness via weak values using two copies of the state}
\label{method}
In this section, we present a technique using weak values to detect entanglement of any two-qubit state, through a single projective measurement, i.e., measurement in a single setting. It was recently shown \cite{tukia_weak}, that by suitable choice of Hamiltonian and post-selective measurement, weak values can be used to determine the concurrence of any {\it pure} two-qubit state. In this paper we  generalize this idea to {\it any} two-qubit state. For this purpose, we consider only two copies of the two-qubit state in consideration.

Now, let us start by considering Alice and Bob share two copies of a two-qubit state $\rho\in\mathcal{P}_+(\mathcal{H}_A\otimes\mathcal{H}_B)$. The most general form of the density matrix of a two-qubit state (mixed or pure) can be expressed in the following form,
\begin{eqnarray}
\label{den_matrix}
\rho=
\left(
\begin{array}{cccc}
 p & u & v & w \\
 u^* & q & x & y \\
 v^* & x^* & r & z \\
 w^* & y^* & z^* & s \\
\end{array}
\right)
\end{eqnarray}
where, $p$, $q$, $r$ and $s$ are real, non-negative numbers summing up to $1$, and $u$, $v$, $w$, $x$, $y$ and $z$ are complex numbers in general; $u^*$ is the complex conjugate of u, etc.  
It should be noted that $\rho$ is Hermitian. In addition to these conditions there is another constraint of positivity of the above matrix, which has to be satisfied by $\rho$ to be a valid density matrix,  
but for our purpose here, we stick to the form given in Eq. (\ref{den_matrix}).

\subsection{The general case}

We first consider the general case, where $p$, $q$, $r$ and $s$ are nonzero. As a result, the determinant of the partially transposed matrix of $\rho$ can be written as,
\begin{eqnarray}
det (\rho^{T_B}) & = & p q r s \Big{(} \frac{u u^* z z^*}{p q r s}-\frac{u v y^* z^*}{p q r s}-\frac{u w^* x z}{p q r s}-\frac{u^* v^* y z}{p q r s} \nonumber \\
&&-\frac{u^* w x^* z^*}{p q r s}+\frac{v v^* y y^*}{p q r s}-\frac{v w^* x^* y}{p q r s}-\frac{v^* w x y^*}{p q r s} \nonumber \\
&&+\frac{w w^* x x^*}{p q r s}+\frac{u v w^*}{p q r}+\frac{u^* v^* w}{p q r}+\frac{u x y^*}{p q s} \nonumber \\
&&+\frac{u^* x^* y}{p q s}-\frac{u u^*}{p q}+\frac{v x^* z^*}{p r s}+\frac{v^* x z}{p r s} \nonumber \\
&&-\frac{v v^*}{p r}-\frac{x x^*}{p s}+\frac{w y^* z^*}{q r s}+\frac{w^* y z}{q r s} \nonumber \\
&&-\frac{w w^*}{q r}-\frac{y y^*}{q s}-\frac{z z^*}{r s}+1 \Big{)}.
\label{det_pt}
\end{eqnarray}
It can be seen that the determinant in Eq. (\ref{det_pt}) is a polynomial of degree 4. In \cite{Grasslpolynomial}, it was shown that an $n$-th degree homogeneous polynomial function of the density matrix elements can be computed as the expectation value of a pair observables, which acts on $n$ copies of the density matrix. This result was later on used by Augusiak et al. \cite{ADH_08} to construct a single observable, acting on four copies of a two-qubit state, to compute the determinant in Eq. (\ref{det_pt}) for witnessing  entanglement.

Our aim is to reduce the number of copies of the state required, and hence reduce the resources required for the process of witnessing. For this purpose we consider the technique of using weak measurement as in \cite{tukia_weak}. Note that in Eq. (\ref{det_sep}), for detecting entanglement of the unknown state $\rho$, 
it is sufficient to know the sign of the determinant in Eq. (\ref{det_pt}). In other words it is enough to find the value of $(1/pqrs)~det{\rho^{T_B}}$. We also found that,  finding values of the following terms (and thereby, their complex conjugates) is sufficient to  determine the value of $(1/pqrs)~det{\rho^{T_B}}$:
\begin{eqnarray}
\frac{u^*}{p}, \frac{u}{q}, \frac{z^*}{r}, \frac{z}{s}, \frac{v^*}{p}, \frac{y^*}{q}, \frac{v}{r}, \frac{y}{s}, \frac{w^*}{p}, \frac{x^*}{q}, \frac{x}{r}, \frac{w}{s}.
\label{generator}
\end{eqnarray}
Out of these 12 terms, it can be easily seen that 9 of them are independent. For example  $\frac{u}{q}$, $\frac{z}{s}$ and $\frac{w^*}{p}$ can be expressed in terms of the 
remaining 9 terms.  
Note that this latter condition does not result in any reduction of copies required for our protocol.

\begin{figure}
  \includegraphics[width= 8.6 cm,height= 4.0 cm]{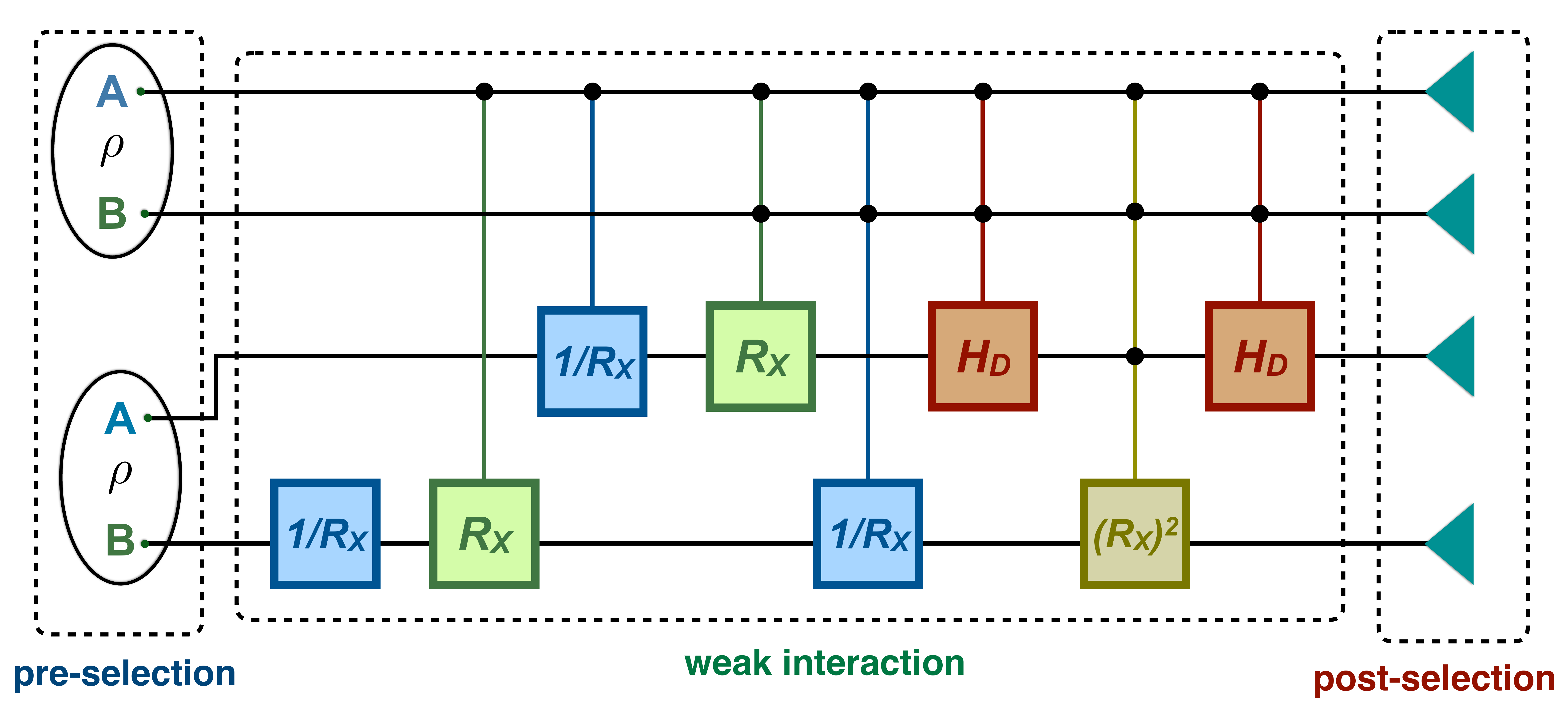}
  \caption{Circuit realization of entanglement detection through weak interaction. Here $R_X=e^{i \epsilon \sigma_x}$ represents rotation of the bloch vector about x-axis through an angle $-2\epsilon$, and $H_D=\ket{0}\bra{+}+\ket{1}\bra{-}$ represents the hadamard operation, where $\ket{\pm}=\frac{1}{\sqrt{2}}\big(\ket{0}\pm\ket{1}\big)$.}
  \label{circuit}
 \end{figure}

We find that each of the terms in Eq. (\ref{generator}) can be seen as a weak value, as in Eq. (\ref{weakvalue}), if we consider two copies of the state i.e. $\rho\otimes\rho\in\mathcal{P}_+((\mathcal{H}_A\otimes\mathcal{H}_B)\otimes(\mathcal{H}_{A}\otimes\mathcal{H}_{B}))$ and choose the Hamiltonian $H\in\mathcal{L}((\mathcal{H}_A\otimes\mathcal{H}_B)\otimes(\mathcal{H}_{A}\otimes\mathcal{H}_{B}))$ in an appropriate form, along with the post-selective measurement in the computational basis i.e., $\{\ket{u_k}\}_{k=1}^{16}=\{\ket{0000},\ket{0001},\dots,\ket{1111}\}$.
It turns out that a suitable form of $H$ is the following,
\begin{eqnarray}
H & = & \ket{00} \bra{00} \otimes H_{1} + \ket{01} \bra{01} \otimes H_{1} \nonumber\\
&& + \ket{10} \bra{10} \otimes H_{2}+ \ket{11} \bra{11} \otimes H_{3}
\label{obs}
\end{eqnarray}
where,
\begin{eqnarray}
H_{1}&=&\openone \otimes \sigma_x \\
H_{2}&=&\sigma_x \otimes \openone \\
H_{3}&=&\sigma_x \otimes \sigma_x 
\label{subobservable}
\end{eqnarray}
with $\sigma_x$ being the usual Pauli matrix along $x$-direction. 
Using the computational basis $\{\ket{u_k}\}_{k=1}^{16}$ and Eqs. (\ref{den_matrix}) and (\ref{obs}), in Eq. (\ref{weakvalue}), we find a list of weak values and the terms of Eq. (\ref{generator}), they correspond to,
\begin{eqnarray}
 \frac{u^*}{p}=\langle H \rangle^{(1)}_{\rho\otimes\rho}~~ &;&~~ \frac{u}{q}=\langle H \rangle^{(2)}_{\rho\otimes\rho}~~ ;~~ \frac{z^*}{r}=\langle H \rangle^{(3)}_{\rho\otimes\rho}~;~~\nonumber\\
  \frac{z}{s}=\langle H \rangle^{(4)}_{\rho\otimes\rho}~~ &;&~~ \frac{v^*}{p}=\langle H \rangle^{(9)}_{\rho\otimes\rho}~ ;~~ \frac{y^*}{q}=\langle H \rangle^{(10)}_{\rho\otimes\rho}~;~~\nonumber\\
   \frac{v}{r}=\langle H \rangle^{(11)}_{\rho\otimes\rho}~~ &;&~~ \frac{y}{s}=\langle H \rangle^{(12)}_{\rho\otimes\rho}~~ ;~~ \frac{w^*}{p}=\langle H \rangle^{(13)}_{\rho\otimes\rho}~;~~\nonumber\\
    \frac{x^*}{q}=\langle H \rangle^{(14)}_{\rho\otimes\rho}~~ &;&~~ \frac{x}{r}=\langle H \rangle^{(15)}_{\rho\otimes\rho}~~ ;~~ \frac{w}{s}=\langle H \rangle^{(16)}_{\rho\otimes\rho}~~
    \label{weaklist}
\end{eqnarray}
Note that four weak values, generated out of the post selection measurement, are redundant. As a result they do not occur in the above equation.
Therefore, it can be easily seen from Eq. (\ref{weaklist}), that our protocol
leads to determination of the sign of the determinant in Eq. (\ref{det_pt}), and as a results it would lead to universal entanglement detection for two-qubit states.
Note that in this protocol, detection of entanglement is made only through a single projective measurement setting, i.e., the post-selective measurement.Using the form of Hamiltonian given in Eq. (\ref{obs}), we find the unitary operator $U$, giving rise to weak interaction is given by,
\begin{eqnarray}
 U&=& \ket{0} \bra{0}\otimes\openone \otimes\openone\otimes e^{-i\epsilon \sigma_x}+\ket{10} \bra{10} \otimes e^{-i\epsilon \sigma_x}\otimes\openone\nonumber\\
&& +\ket{11} \bra{11} \otimes e^{-i\epsilon \sigma_x\otimes\sigma_x}
\end{eqnarray}
In the above form, we can write $e^{-i\epsilon\sigma_x\otimes\sigma_x}=\ket{+}\bra{+}\otimes e^{-i\epsilon\sigma_x}+\ket{-}\bra{-}\otimes e^{i\epsilon\sigma_x}$. Note that, this represents a conditional unitary operation, conditioned on $\{\ket{+},\ket{-}\}$ states. As a result we use Hadamard gate $H_D$, which flips states $\{\ket{+},\ket{-\}}\leftrightarrow\{\ket{0},\ket{1}\}$, to achieve the circuit realization of our protocol, as given in Fig. \ref{circuit}.
Also note that, the decomposition of $H$ in Eq.(\ref{obs}) is not unique, as it  can also be chosen in any form 
where $H_{1}$, $H_{2}$ and $H_{3}$ 
resides in any of the diagonal blocks of the $16 \times 16$ matrix $H$.  Moreover, as mentioned earlier, there are 9 independent terms in Eq. (\ref{generator}), which leads to 9 independent linear equations. Along with these equations, the constraint $p+q+r+s=1$, gives the exact solution for all the unknown quantities in the density matrix $\rho$, and hence our protocol results in complete identification of the two-qubit state i.e state tomography. Henceforth, any standard method for finding the amount of entanglement like negativity   \cite{negativity,negativity2} or concurrence \cite{coffman} can be employed, and one can calculate how much entangled the state is. In particular, the following quantity, which can be easily obtained from our protocol, can also used to estimate the amount of entanglement present in the state :
\begin{equation}
 E(\rho)=max\{0,-det(\rho^{T_B})\}.
\end{equation}
Thus we see our protocol not only serves as a technique to detect arbitrary two-qubit entangled state, but also as a protocol to measure entanglement.

\subsection{Special cases}

Let us now consider the special cases where at least one of the diagonal elements of $\rho$ is zero. This scenario physically means receiving no signal on the pointer, for the corresponding measurement outcome, before the weak interaction is switched on. For example, if $p$ is $0$, no signal is received for outcome $\ket{0000}$, and similarly for $q,r$ or $s$ we check if no signal is received for the outcomes $\ket{0101},\ket{1010}$ or $\ket{1111}$, respectively.
 We will show here that even for this case the same protocol, as described in Fig. \ref{circuit}, works. 
We now consider each case individually,
\subsubsection{Case I}
When $p=0$, positivity of $\rho$ demands $u$, $v$ and $w$ must also be $0$. Similarly, when $s=0$ we must have $w=y=z=0$. As a result, in both of these cases, we have $det (\rho^{T_B})= - x x^* q r$. In both of these cases, we first check if $q$ or $r$ is zero or not. If either of them is zero, we conclude $\rho$ is separable. If not, we check if the weak value $\langle H \rangle^{(14)}_{\rho\otimes\rho}$ i e. $x^*/q$, is zero. If it is, then $\rho$ is separable, otherwise it is entangled.

\subsubsection{Case II}
Similarly, when $q=0$ or $r=0$ we have $u=x=y=0$ or $v=x=z=0$, respectively. In both the cases we get $det (\rho^{T_B})= - w w^* p s$. Following this, in a similar way as above, we check the values of $p$ or $s$ and subsequently, the weak value $\langle H \rangle^{(16)}_{\rho\otimes\rho}$ i e. $w/s$, to determine if $\rho$ is entangled or not.

\subsection{Implementing the protocol through Local Operations}
\label{local impementation}
In this  section, we show that if we restrict the two-qubit state to be {\it pure} states only, we can realize the weak interaction through local operations on each of the qubits. Consider $\rho=\ket{\Psi}\bra{\Psi}$, where $\ket{\Psi}=a\ket{00}+b\ket{01}+c\ket{10}+d\ket{11}$. It can be easily seen that $\rho$ is separable if and only if $ad-bc=0$. In the notations of Eq. (\ref{den_matrix}), we find $p=|a|^2,q=|b|^2,r=|c|^2,s=|d|^2,u=ab^*$ and $z=cd^*$. Therefore, for this case we modify our protocol, and choose the weak Hamiltonian in Eq (\ref{weakvalue}) to be of the form, 
\begin{eqnarray}
 H'=\openone\otimes\openone\otimes\openone\otimes\sigma_x
 \label{localhamil}
\end{eqnarray}
It can be easily checked that the unitary operator corresponding to this Hamiltonian acts locally on all of the four qubits. Now, in the same way as in the previous section, we first check if either of $p,q,r$ or $s$ is zero i.e by checking if signal is received for outcomes $\ket{0000},\ket{0101},\ket{1010}$ or $\ket{1111}$. If one, or more of $p,q,r$ and $s$ are zero,
it would imply the corresponding terms among $a,b,c$ and $d$ are also zero. Using these values we can easily check if $ad-bc=0$. If none of $p,q,r$ or $s$
are zero, we check if the weak values $\langle H' \rangle^{(2)}_{\rho\otimes\rho}=\langle H' \rangle^{(4)}_{\rho\otimes\rho}$, i e. if $u/q=z/s$. If the equality holds then $\rho$ is separable, otherwise it is entangled. 

\section{robustness of the protocol}
\label{robust}

In real life experiments, errors are bound to occur. Here we show that, our protocol is robust against errors arising from inappropriate choice of weak interaction. Consider a situation, where an erroneous Hamiltonian of the form of $ H_e$ is chosen in place of the correct Hamiltonian $H$, where $||H-H_e||_1\leq\delta$. Note here, $||A||_1=tr\sqrt{A^{\dagger}A}$ represents the {\it trace norm} of a matrix $A$. As a result, the error occuring in the weak values are given by,
\begin{eqnarray}
\Delta_k &=&|\langle H \rangle^{(k)}_{\rho\otimes\rho}-\langle H_e \rangle^{(k)}_{\rho\otimes\rho}|=\frac{|\bra{u_k}\rho\otimes\rho (H-H_e)\ket{u_k}|}{\bra{u_k}\rho\otimes\rho\ket{u_k}}\nonumber\\
 &\leq&\frac{|\bra{u_k}\rho\otimes\rho (H-H_e)\ket{u_k}|}{m}
 \label{robust1}
\end{eqnarray}
where $m$ is the minimum of $\big\{\{p,q,r,s\}\times\{p,q,r,s\}\big\}$ \cite{crossprod} and is always positive. Note that, in obtaining the above inequality we used the fact $\bra{u_k}\rho\otimes\rho\ket{u_k}\in\big\{\{p,q,r,s\}\times\{p,q,r,s\}\big\}$ and it can also be  seen that in our protocol, the weak value for $k^{th}$ outcome is only measured when $\bra{u_k}\rho\otimes\rho\ket{u_k}\neq0$. As a result, the denominator never vanishes in Eq. (\ref{robust1}). Now, consider the eigenvalue decomposition $H-H_e=\sum_{i}\lambda_i\ket{i}\bra{i}$, where $\{\ket{i}\}_i$ forms an orthonormal basis in $\mathcal{H}_A\otimes\mathcal{H}_B\otimes\mathcal{H}_A\otimes\mathcal{H}_B$, and also note that $||H-H_e||_1=\sum_i|\lambda_i|$. As a result,
\begin{eqnarray}
 \Delta_k &\leq&\frac{|\sum_i\lambda_i\bra{u_k}\rho\otimes\rho \ket{i}\langle{i}|{u_k}\rangle|}{m}\nonumber\\
 &\leq&\frac{1}{m}\sum_i|\lambda_i|~|\bra{u_k}\rho\otimes\rho \ket{i}\langle{i}|{u_k}\rangle|.
\end{eqnarray}
Since $0\leq\rho\leq\openone$, it can be easily seen that $|\bra{u_k}\rho\otimes\rho \ket{i}\langle{i}|{u_k}\rangle|\leq 1$. Thus we have,
\begin{eqnarray}
 \Delta_k \leq\frac{1}{m}\sum_i|\lambda_i|\leq \frac{||H-H_e||_1}{m}\leq\frac{\delta}{m }
\end{eqnarray}
Thus we see, our protocol is robust to errors arising from inappropriate choice of weak interaction.\\

\section{Conclusions}
\label{conclusion}
In this paper, we have proposed a universal entanglement detection protocol for two-qubit quantum states. We consider the most general form of density matrix for two-qubit states, and  show that it is enough to have just two copies of the given state to identify if it is entangled or not. Our formulation is based on  the determinant based separability criterion and the idea of weak values. Previously in \cite{ADH_08}, it was  demonstrated that one can universally detect entanglement in two-qubit systems using four copies of the state. Our protocol therefore, leads to a clear advantage in terms of resource, as in our case it is sufficient to have just two copies of the state. Our protocol requires only a single projective measurement setting in the computational basis for the purpose of post selection in weak measurement. It is interesting to note that in our protocol the number of copies required for entanglement detection may be further reduced if some partial information about the state is known. Moreover, we have also shown that the procedure of identification is achievable by local operations, if the state in consideration is a pure state. Further, we have shown that the protocol is robust against error arising during application of the weak interaction. 

Before concluding, it may be noted that though our scheme reduces the number of measurement settings
compared to the universal entanglement witnessing scheme of \cite{ADH_08} that requires four copies of the state at a time, this advantage comes at the expense of joint unitary actions on two copies of the state (for arbitrary mixed states). Further work involving quantitative comparison of resources used in our scheme and that employed in other schemes such as in \cite{ADH_08} would be needed to obtain
a clear idea of practical viability. In this context, one may need to compare the energy cost of creating correlations \cite{energycorrelation} with the energy cost of doing measurements \cite{energymeasurement1,energymeasurement2} used in the various protocols. Finally, we note that if a similar determinant based criterion for identification of certain class of states is available for higher dimensions, we expect a similar detection protocol such as ours to work therein.

\begin{acknowledgments}
 SC would like to acknowledge Arun K Pati for having useful discussions on weak values and weak measurement. 
\end{acknowledgments}

\bibliography{ref_weak_1}

\begin{thebibliography}{70}%
\makeatletter
\providecommand \@ifxundefined [1]{%
 \@ifx{#1\undefined}
}%
\providecommand \@ifnum [1]{%
 \ifnum #1\expandafter \@firstoftwo
 \else \expandafter \@secondoftwo
 \fi
}%
\providecommand \@ifx [1]{%
 \ifx #1\expandafter \@firstoftwo
 \else \expandafter \@secondoftwo
 \fi
}%
\providecommand \natexlab [1]{#1}%
\providecommand \enquote  [1]{``#1''}%
\providecommand \bibnamefont  [1]{#1}%
\providecommand \bibfnamefont [1]{#1}%
\providecommand \citenamefont [1]{#1}%
\providecommand \href@noop [0]{\@secondoftwo}%
\providecommand \href [0]{\begingroup \@sanitize@url \@href}%
\providecommand \@href[1]{\@@startlink{#1}\@@href}%
\providecommand \@@href[1]{\endgroup#1\@@endlink}%
\providecommand \@sanitize@url [0]{\catcode `\\12\catcode `\$12\catcode
  `\&12\catcode `\#12\catcode `\^12\catcode `\_12\catcode `\%12\relax}%
\providecommand \@@startlink[1]{}%
\providecommand \@@endlink[0]{}%
\providecommand \url  [0]{\begingroup\@sanitize@url \@url }%
\providecommand \@url [1]{\endgroup\@href {#1}{\urlprefix }}%
\providecommand \urlprefix  [0]{URL }%
\providecommand \Eprint [0]{\href }%
\providecommand \doibase [0]{http://dx.doi.org/}%
\providecommand \selectlanguage [0]{\@gobble}%
\providecommand \bibinfo  [0]{\@secondoftwo}%
\providecommand \bibfield  [0]{\@secondoftwo}%
\providecommand \translation [1]{[#1]}%
\providecommand \BibitemOpen [0]{}%
\providecommand \bibitemStop [0]{}%
\providecommand \bibitemNoStop [0]{.\EOS\space}%
\providecommand \EOS [0]{\spacefactor3000\relax}%
\providecommand \BibitemShut  [1]{\csname bibitem#1\endcsname}%
\let\auto@bib@innerbib\@empty
\bibitem [{\citenamefont {Einstein}\ \emph {et~al.}(1935)\citenamefont
  {Einstein}, \citenamefont {Podolsky},\ and\ \citenamefont {Rosen}}]{EPR_35}%
  \BibitemOpen
  \bibfield  {author} {\bibinfo {author} {\bibfnamefont {A.}~\bibnamefont
  {Einstein}}, \bibinfo {author} {\bibfnamefont {B.}~\bibnamefont {Podolsky}},
  \ and\ \bibinfo {author} {\bibfnamefont {N.}~\bibnamefont {Rosen}},\ }\href
  {\doibase 10.1103/PhysRev.47.777} {\bibfield  {journal} {\bibinfo  {journal}
  {Phys. Rev.}\ }\textbf {\bibinfo {volume} {47}},\ \bibinfo {pages} {777}
  (\bibinfo {year} {1935})}\BibitemShut {NoStop}%
\bibitem [{\citenamefont {Schrodinger}(1935)}]{S_35}%
  \BibitemOpen
  \bibfield  {author} {\bibinfo {author} {\bibfnamefont {E.}~\bibnamefont
  {Schrodinger}},\ }\href {\doibase 10.1017/S0305004100013554} {\bibfield
  {journal} {\bibinfo  {journal} {Mathematical Proceedings of the Cambridge
  Philosophical Society}\ }\textbf {\bibinfo {volume} {31}},\ \bibinfo {pages}
  {555} (\bibinfo {year} {1935})}\BibitemShut {NoStop}%
\bibitem [{\citenamefont {Bell}(1964)}]{B_64}%
  \BibitemOpen
  \bibfield  {author} {\bibinfo {author} {\bibfnamefont {J.~S.}\ \bibnamefont
  {Bell}},\ }\href {\doibase 10.1103/PhysicsPhysiqueFizika.1.195} {\bibfield
  {journal} {\bibinfo  {journal} {Physics Physique Fizika}\ }\textbf {\bibinfo
  {volume} {1}},\ \bibinfo {pages} {195} (\bibinfo {year} {1964})}\BibitemShut
  {NoStop}%
\bibitem [{\citenamefont {Wiseman}\ \emph {et~al.}(2007)\citenamefont
  {Wiseman}, \citenamefont {Jones},\ and\ \citenamefont {Doherty}}]{WJD_07}%
  \BibitemOpen
  \bibfield  {author} {\bibinfo {author} {\bibfnamefont {H.~M.}\ \bibnamefont
  {Wiseman}}, \bibinfo {author} {\bibfnamefont {S.~J.}\ \bibnamefont {Jones}},
  \ and\ \bibinfo {author} {\bibfnamefont {A.~C.}\ \bibnamefont {Doherty}},\
  }\href {\doibase 10.1103/PhysRevLett.98.140402} {\bibfield  {journal}
  {\bibinfo  {journal} {Phys. Rev. Lett.}\ }\textbf {\bibinfo {volume} {98}},\
  \bibinfo {pages} {140402} (\bibinfo {year} {2007})}\BibitemShut {NoStop}%
\bibitem [{\citenamefont {Jones}\ \emph {et~al.}(2007)\citenamefont {Jones},
  \citenamefont {Wiseman},\ and\ \citenamefont {Doherty}}]{JWD_07}%
  \BibitemOpen
  \bibfield  {author} {\bibinfo {author} {\bibfnamefont {S.~J.}\ \bibnamefont
  {Jones}}, \bibinfo {author} {\bibfnamefont {H.~M.}\ \bibnamefont {Wiseman}},
  \ and\ \bibinfo {author} {\bibfnamefont {A.~C.}\ \bibnamefont {Doherty}},\
  }\href {\doibase 10.1103/PhysRevA.76.052116} {\bibfield  {journal} {\bibinfo
  {journal} {Phys. Rev. A}\ }\textbf {\bibinfo {volume} {76}},\ \bibinfo
  {pages} {052116} (\bibinfo {year} {2007})}\BibitemShut {NoStop}%
\bibitem [{\citenamefont {Horodecki}\ \emph {et~al.}(2009)\citenamefont
  {Horodecki}, \citenamefont {Horodecki}, \citenamefont {Horodecki},\ and\
  \citenamefont {Horodecki}}]{enatnglement_review}%
  \BibitemOpen
  \bibfield  {author} {\bibinfo {author} {\bibfnamefont {R.}~\bibnamefont
  {Horodecki}}, \bibinfo {author} {\bibfnamefont {P.}~\bibnamefont
  {Horodecki}}, \bibinfo {author} {\bibfnamefont {M.}~\bibnamefont
  {Horodecki}}, \ and\ \bibinfo {author} {\bibfnamefont {K.}~\bibnamefont
  {Horodecki}},\ }\href {\doibase 10.1103/RevModPhys.81.865} {\bibfield
  {journal} {\bibinfo  {journal} {Rev. Mod. Phys.}\ }\textbf {\bibinfo {volume}
  {81}},\ \bibinfo {pages} {865} (\bibinfo {year} {2009})}\BibitemShut
  {NoStop}%
\bibitem [{\citenamefont {Bennett}\ \emph {et~al.}(1993)\citenamefont
  {Bennett}, \citenamefont {Brassard}, \citenamefont {Cr\'epeau}, \citenamefont
  {Jozsa}, \citenamefont {Peres},\ and\ \citenamefont
  {Wootters}}]{teleportation}%
  \BibitemOpen
  \bibfield  {author} {\bibinfo {author} {\bibfnamefont {C.~H.}\ \bibnamefont
  {Bennett}}, \bibinfo {author} {\bibfnamefont {G.}~\bibnamefont {Brassard}},
  \bibinfo {author} {\bibfnamefont {C.}~\bibnamefont {Cr\'epeau}}, \bibinfo
  {author} {\bibfnamefont {R.}~\bibnamefont {Jozsa}}, \bibinfo {author}
  {\bibfnamefont {A.}~\bibnamefont {Peres}}, \ and\ \bibinfo {author}
  {\bibfnamefont {W.~K.}\ \bibnamefont {Wootters}},\ }\href {\doibase
  10.1103/PhysRevLett.70.1895} {\bibfield  {journal} {\bibinfo  {journal}
  {Phys. Rev. Lett.}\ }\textbf {\bibinfo {volume} {70}},\ \bibinfo {pages}
  {1895} (\bibinfo {year} {1993})}\BibitemShut {NoStop}%
\bibitem [{\citenamefont {Bennett}\ and\ \citenamefont
  {Wiesner}(1992)}]{dense}%
  \BibitemOpen
  \bibfield  {author} {\bibinfo {author} {\bibfnamefont {C.~H.}\ \bibnamefont
  {Bennett}}\ and\ \bibinfo {author} {\bibfnamefont {S.~J.}\ \bibnamefont
  {Wiesner}},\ }\href {\doibase 10.1103/PhysRevLett.69.2881} {\bibfield
  {journal} {\bibinfo  {journal} {Phys. Rev. Lett.}\ }\textbf {\bibinfo
  {volume} {69}},\ \bibinfo {pages} {2881} (\bibinfo {year}
  {1992})}\BibitemShut {NoStop}%
\bibitem [{\citenamefont {Branciard}\ \emph {et~al.}(2012)\citenamefont
  {Branciard}, \citenamefont {Cavalcanti}, \citenamefont {Walborn},
  \citenamefont {Scarani},\ and\ \citenamefont {Wiseman}}]{1sqkd}%
  \BibitemOpen
  \bibfield  {author} {\bibinfo {author} {\bibfnamefont {C.}~\bibnamefont
  {Branciard}}, \bibinfo {author} {\bibfnamefont {E.~G.}\ \bibnamefont
  {Cavalcanti}}, \bibinfo {author} {\bibfnamefont {S.~P.}\ \bibnamefont
  {Walborn}}, \bibinfo {author} {\bibfnamefont {V.}~\bibnamefont {Scarani}}, \
  and\ \bibinfo {author} {\bibfnamefont {H.~M.}\ \bibnamefont {Wiseman}},\
  }\href {\doibase 10.1103/PhysRevA.85.010301} {\bibfield  {journal} {\bibinfo
  {journal} {Phys. Rev. A}\ }\textbf {\bibinfo {volume} {85}},\ \bibinfo
  {pages} {010301} (\bibinfo {year} {2012})}\BibitemShut {NoStop}%
\bibitem [{\citenamefont {Ac\'{\i}n}\ \emph {et~al.}(2012)\citenamefont
  {Ac\'{\i}n}, \citenamefont {Massar},\ and\ \citenamefont
  {Pironio}}]{randomness}%
  \BibitemOpen
  \bibfield  {author} {\bibinfo {author} {\bibfnamefont {A.}~\bibnamefont
  {Ac\'{\i}n}}, \bibinfo {author} {\bibfnamefont {S.}~\bibnamefont {Massar}}, \
  and\ \bibinfo {author} {\bibfnamefont {S.}~\bibnamefont {Pironio}},\ }\href
  {\doibase 10.1103/PhysRevLett.108.100402} {\bibfield  {journal} {\bibinfo
  {journal} {Phys. Rev. Lett.}\ }\textbf {\bibinfo {volume} {108}},\ \bibinfo
  {pages} {100402} (\bibinfo {year} {2012})}\BibitemShut {NoStop}%
\bibitem [{\citenamefont {Pironio}\ \emph {et~al.}(2010)\citenamefont
  {Pironio}, \citenamefont {Ac{\'\i}n}, \citenamefont {Massar}, \citenamefont
  {de~La~Giroday}, \citenamefont {Matsukevich}, \citenamefont {Maunz},
  \citenamefont {Olmschenk}, \citenamefont {Hayes}, \citenamefont {Luo},
  \citenamefont {Manning} \emph {et~al.}}]{pironio2010random}%
  \BibitemOpen
  \bibfield  {author} {\bibinfo {author} {\bibfnamefont {S.}~\bibnamefont
  {Pironio}}, \bibinfo {author} {\bibfnamefont {A.}~\bibnamefont {Ac{\'\i}n}},
  \bibinfo {author} {\bibfnamefont {S.}~\bibnamefont {Massar}}, \bibinfo
  {author} {\bibfnamefont {A.~B.}\ \bibnamefont {de~La~Giroday}}, \bibinfo
  {author} {\bibfnamefont {D.~N.}\ \bibnamefont {Matsukevich}}, \bibinfo
  {author} {\bibfnamefont {P.}~\bibnamefont {Maunz}}, \bibinfo {author}
  {\bibfnamefont {S.}~\bibnamefont {Olmschenk}}, \bibinfo {author}
  {\bibfnamefont {D.}~\bibnamefont {Hayes}}, \bibinfo {author} {\bibfnamefont
  {L.}~\bibnamefont {Luo}}, \bibinfo {author} {\bibfnamefont {T.~A.}\
  \bibnamefont {Manning}},  \emph {et~al.},\ }\href@noop {} {\bibfield
  {journal} {\bibinfo  {journal} {Nature}\ }\textbf {\bibinfo {volume} {464}},\
  \bibinfo {pages} {1021} (\bibinfo {year} {2010})}\BibitemShut {NoStop}%
\bibitem [{\citenamefont {Peres}(1996)}]{peres_sep}%
  \BibitemOpen
  \bibfield  {author} {\bibinfo {author} {\bibfnamefont {A.}~\bibnamefont
  {Peres}},\ }\href {\doibase 10.1103/PhysRevLett.77.1413} {\bibfield
  {journal} {\bibinfo  {journal} {Phys. Rev. Lett.}\ }\textbf {\bibinfo
  {volume} {77}},\ \bibinfo {pages} {1413} (\bibinfo {year}
  {1996})}\BibitemShut {NoStop}%
\bibitem [{\citenamefont {Horodecki}\ \emph {et~al.}(1996)\citenamefont
  {Horodecki}, \citenamefont {Horodecki},\ and\ \citenamefont
  {Horodecki}}]{HORODECKI19961}%
  \BibitemOpen
  \bibfield  {author} {\bibinfo {author} {\bibfnamefont {M.}~\bibnamefont
  {Horodecki}}, \bibinfo {author} {\bibfnamefont {P.}~\bibnamefont
  {Horodecki}}, \ and\ \bibinfo {author} {\bibfnamefont {R.}~\bibnamefont
  {Horodecki}},\ }\href {\doibase
  https://doi.org/10.1016/S0375-9601(96)00706-2} {\bibfield  {journal}
  {\bibinfo  {journal} {Physics Letters A}\ }\textbf {\bibinfo {volume}
  {223}},\ \bibinfo {pages} {1 } (\bibinfo {year} {1996})}\BibitemShut
  {NoStop}%
\bibitem [{\citenamefont {Vidal}\ and\ \citenamefont
  {Werner}(2002)}]{negativity}%
  \BibitemOpen
  \bibfield  {author} {\bibinfo {author} {\bibfnamefont {G.}~\bibnamefont
  {Vidal}}\ and\ \bibinfo {author} {\bibfnamefont {R.~F.}\ \bibnamefont
  {Werner}},\ }\href {\doibase 10.1103/PhysRevA.65.032314} {\bibfield
  {journal} {\bibinfo  {journal} {Phys. Rev. A}\ }\textbf {\bibinfo {volume}
  {65}},\ \bibinfo {pages} {032314} (\bibinfo {year} {2002})}\BibitemShut
  {NoStop}%
\bibitem [{\citenamefont {Audenaert}\ \emph {et~al.}(2003)\citenamefont
  {Audenaert}, \citenamefont {Plenio},\ and\ \citenamefont
  {Eisert}}]{log_negativity}%
  \BibitemOpen
  \bibfield  {author} {\bibinfo {author} {\bibfnamefont {K.}~\bibnamefont
  {Audenaert}}, \bibinfo {author} {\bibfnamefont {M.~B.}\ \bibnamefont
  {Plenio}}, \ and\ \bibinfo {author} {\bibfnamefont {J.}~\bibnamefont
  {Eisert}},\ }\href {\doibase 10.1103/PhysRevLett.90.027901} {\bibfield
  {journal} {\bibinfo  {journal} {Phys. Rev. Lett.}\ }\textbf {\bibinfo
  {volume} {90}},\ \bibinfo {pages} {027901} (\bibinfo {year}
  {2003})}\BibitemShut {NoStop}%
\bibitem [{\citenamefont {Reh{\'a}cek}\ and\ \citenamefont
  {Paris}(2004)}]{state_estimation_book}%
  \BibitemOpen
  \bibfield  {author} {\bibinfo {author} {\bibfnamefont {J.}~\bibnamefont
  {Reh{\'a}cek}}\ and\ \bibinfo {author} {\bibfnamefont {M.}~\bibnamefont
  {Paris}},\ }\href@noop {} {\bibfield  {journal} {\bibinfo  {journal} {Lect.
  Notes Phys}\ }\textbf {\bibinfo {volume} {649}} (\bibinfo {year}
  {2004})}\BibitemShut {NoStop}%
\bibitem [{\citenamefont {Terhal}(2000)}]{T_00}%
  \BibitemOpen
  \bibfield  {author} {\bibinfo {author} {\bibfnamefont {B.~M.}\ \bibnamefont
  {Terhal}},\ }\href@noop {} {\bibfield  {journal} {\bibinfo  {journal}
  {Physics Letters A}\ }\textbf {\bibinfo {volume} {271}},\ \bibinfo {pages}
  {319} (\bibinfo {year} {2000})}\BibitemShut {NoStop}%
\bibitem [{\citenamefont {Lewenstein}\ \emph {et~al.}(2000)\citenamefont
  {Lewenstein}, \citenamefont {Kraus}, \citenamefont {Cirac},\ and\
  \citenamefont {Horodecki}}]{LKCH_00}%
  \BibitemOpen
  \bibfield  {author} {\bibinfo {author} {\bibfnamefont {M.}~\bibnamefont
  {Lewenstein}}, \bibinfo {author} {\bibfnamefont {B.}~\bibnamefont {Kraus}},
  \bibinfo {author} {\bibfnamefont {J.~I.}\ \bibnamefont {Cirac}}, \ and\
  \bibinfo {author} {\bibfnamefont {P.}~\bibnamefont {Horodecki}},\ }\href
  {\doibase 10.1103/PhysRevA.62.052310} {\bibfield  {journal} {\bibinfo
  {journal} {Phys. Rev. A}\ }\textbf {\bibinfo {volume} {62}},\ \bibinfo
  {pages} {052310} (\bibinfo {year} {2000})}\BibitemShut {NoStop}%
\bibitem [{\citenamefont {Bru{\ss}}\ \emph {et~al.}(2002)\citenamefont
  {Bru{\ss}}, \citenamefont {Cirac}, \citenamefont {Horodecki}, \citenamefont
  {Hulpke}, \citenamefont {Kraus}, \citenamefont {Lewenstein},\ and\
  \citenamefont {Sanpera}}]{bruss2002}%
  \BibitemOpen
  \bibfield  {author} {\bibinfo {author} {\bibfnamefont {D.}~\bibnamefont
  {Bru{\ss}}}, \bibinfo {author} {\bibfnamefont {J.~I.}\ \bibnamefont {Cirac}},
  \bibinfo {author} {\bibfnamefont {P.}~\bibnamefont {Horodecki}}, \bibinfo
  {author} {\bibfnamefont {F.}~\bibnamefont {Hulpke}}, \bibinfo {author}
  {\bibfnamefont {B.}~\bibnamefont {Kraus}}, \bibinfo {author} {\bibfnamefont
  {M.}~\bibnamefont {Lewenstein}}, \ and\ \bibinfo {author} {\bibfnamefont
  {A.}~\bibnamefont {Sanpera}},\ }\href@noop {} {\bibfield  {journal} {\bibinfo
   {journal} {Journal of Modern Optics}\ }\textbf {\bibinfo {volume} {49}},\
  \bibinfo {pages} {1399} (\bibinfo {year} {2002})}\BibitemShut {NoStop}%
\bibitem [{\citenamefont {Guo}\ \emph {et~al.}(2018)\citenamefont {Guo},
  \citenamefont {Hu}, \citenamefont {Liu}, \citenamefont {Huang}, \citenamefont
  {Li},\ and\ \citenamefont {Guo}}]{experimental_witness}%
  \BibitemOpen
  \bibfield  {author} {\bibinfo {author} {\bibfnamefont {Y.}~\bibnamefont
  {Guo}}, \bibinfo {author} {\bibfnamefont {X.-M.}\ \bibnamefont {Hu}},
  \bibinfo {author} {\bibfnamefont {B.-H.}\ \bibnamefont {Liu}}, \bibinfo
  {author} {\bibfnamefont {Y.-F.}\ \bibnamefont {Huang}}, \bibinfo {author}
  {\bibfnamefont {C.-F.}\ \bibnamefont {Li}}, \ and\ \bibinfo {author}
  {\bibfnamefont {G.-C.}\ \bibnamefont {Guo}},\ }\href {\doibase
  10.1103/PhysRevA.97.062309} {\bibfield  {journal} {\bibinfo  {journal} {Phys.
  Rev. A}\ }\textbf {\bibinfo {volume} {97}},\ \bibinfo {pages} {062309}
  (\bibinfo {year} {2018})}\BibitemShut {NoStop}%
\bibitem [{\citenamefont {G{\"u}hne}\ \emph {et~al.}(2003)\citenamefont
  {G{\"u}hne}, \citenamefont {Hyllus}, \citenamefont {Bru{\ss}}, \citenamefont
  {Ekert}, \citenamefont {Lewenstein}, \citenamefont {Macchiavello},\ and\
  \citenamefont {Sanpera}}]{guhne2003experimental}%
  \BibitemOpen
  \bibfield  {author} {\bibinfo {author} {\bibfnamefont {O.}~\bibnamefont
  {G{\"u}hne}}, \bibinfo {author} {\bibfnamefont {P.}~\bibnamefont {Hyllus}},
  \bibinfo {author} {\bibfnamefont {D.}~\bibnamefont {Bru{\ss}}}, \bibinfo
  {author} {\bibfnamefont {A.}~\bibnamefont {Ekert}}, \bibinfo {author}
  {\bibfnamefont {M.}~\bibnamefont {Lewenstein}}, \bibinfo {author}
  {\bibfnamefont {C.}~\bibnamefont {Macchiavello}}, \ and\ \bibinfo {author}
  {\bibfnamefont {A.}~\bibnamefont {Sanpera}},\ }\href@noop {} {\bibfield
  {journal} {\bibinfo  {journal} {Journal of Modern Optics}\ }\textbf {\bibinfo
  {volume} {50}},\ \bibinfo {pages} {1079} (\bibinfo {year}
  {2003})}\BibitemShut {NoStop}%
\bibitem [{\citenamefont {McKague}\ \emph {et~al.}(2012)\citenamefont
  {McKague}, \citenamefont {Yang},\ and\ \citenamefont {Scarani}}]{MYS12}%
  \BibitemOpen
  \bibfield  {author} {\bibinfo {author} {\bibfnamefont {M.}~\bibnamefont
  {McKague}}, \bibinfo {author} {\bibfnamefont {T.~H.}\ \bibnamefont {Yang}}, \
  and\ \bibinfo {author} {\bibfnamefont {V.}~\bibnamefont {Scarani}},\ }\href
  {http://stacks.iop.org/1751-8121/45/i=45/a=455304} {\bibfield  {journal}
  {\bibinfo  {journal} {Journal of Physics A: Mathematical and Theoretical}\
  }\textbf {\bibinfo {volume} {45}},\ \bibinfo {pages} {455304} (\bibinfo
  {year} {2012})}\BibitemShut {NoStop}%
\bibitem [{\citenamefont {Yang}\ and\ \citenamefont
  {Navascu\'es}(2013)}]{YN13}%
  \BibitemOpen
  \bibfield  {author} {\bibinfo {author} {\bibfnamefont {T.~H.}\ \bibnamefont
  {Yang}}\ and\ \bibinfo {author} {\bibfnamefont {M.}~\bibnamefont
  {Navascu\'es}},\ }\href {\doibase 10.1103/PhysRevA.87.050102} {\bibfield
  {journal} {\bibinfo  {journal} {Phys. Rev. A}\ }\textbf {\bibinfo {volume}
  {87}},\ \bibinfo {pages} {050102} (\bibinfo {year} {2013})}\BibitemShut
  {NoStop}%
\bibitem [{\citenamefont {Supic}\ and\ \citenamefont {Hoban}(2016)}]{SH16}%
  \BibitemOpen
  \bibfield  {author} {\bibinfo {author} {\bibfnamefont {I.}~\bibnamefont
  {Supic}}\ and\ \bibinfo {author} {\bibfnamefont {M.~J.}\ \bibnamefont
  {Hoban}},\ }\href {http://stacks.iop.org/1367-2630/18/i=7/a=075006}
  {\bibfield  {journal} {\bibinfo  {journal} {New Journal of Physics}\ }\textbf
  {\bibinfo {volume} {18}},\ \bibinfo {pages} {075006} (\bibinfo {year}
  {2016})}\BibitemShut {NoStop}%
\bibitem [{\citenamefont {Goswami}\ \emph {et~al.}(2018)\citenamefont
  {Goswami}, \citenamefont {Bhattacharya}, \citenamefont {Das}, \citenamefont
  {Sasmal}, \citenamefont {Jebaratnam},\ and\ \citenamefont
  {Majumdar}}]{GBDSJM_18}%
  \BibitemOpen
  \bibfield  {author} {\bibinfo {author} {\bibfnamefont {S.}~\bibnamefont
  {Goswami}}, \bibinfo {author} {\bibfnamefont {B.}~\bibnamefont
  {Bhattacharya}}, \bibinfo {author} {\bibfnamefont {D.}~\bibnamefont {Das}},
  \bibinfo {author} {\bibfnamefont {S.}~\bibnamefont {Sasmal}}, \bibinfo
  {author} {\bibfnamefont {C.}~\bibnamefont {Jebaratnam}}, \ and\ \bibinfo
  {author} {\bibfnamefont {A.~S.}\ \bibnamefont {Majumdar}},\ }\href {\doibase
  10.1103/PhysRevA.98.022311} {\bibfield  {journal} {\bibinfo  {journal} {Phys.
  Rev. A}\ }\textbf {\bibinfo {volume} {98}},\ \bibinfo {pages} {022311}
  (\bibinfo {year} {2018})}\BibitemShut {NoStop}%
\bibitem [{\citenamefont {G{\"u}hne}\ and\ \citenamefont
  {T{\'o}th}(2009)}]{guhne2009entanglement}%
  \BibitemOpen
  \bibfield  {author} {\bibinfo {author} {\bibfnamefont {O.}~\bibnamefont
  {G{\"u}hne}}\ and\ \bibinfo {author} {\bibfnamefont {G.}~\bibnamefont
  {T{\'o}th}},\ }\href@noop {} {\bibfield  {journal} {\bibinfo  {journal}
  {Physics Reports}\ }\textbf {\bibinfo {volume} {474}},\ \bibinfo {pages} {1}
  (\bibinfo {year} {2009})}\BibitemShut {NoStop}%
\bibitem [{\citenamefont {Zhou}\ and\ \citenamefont
  {Sheng}(2014)}]{recenttwoqubit3}%
  \BibitemOpen
  \bibfield  {author} {\bibinfo {author} {\bibfnamefont {L.}~\bibnamefont
  {Zhou}}\ and\ \bibinfo {author} {\bibfnamefont {Y.-B.}\ \bibnamefont
  {Sheng}},\ }\href {\doibase 10.1103/PhysRevA.90.024301} {\bibfield  {journal}
  {\bibinfo  {journal} {Phys. Rev. A}\ }\textbf {\bibinfo {volume} {90}},\
  \bibinfo {pages} {024301} (\bibinfo {year} {2014})}\BibitemShut {NoStop}%
\bibitem [{\citenamefont {Rafiee}\ \emph {et~al.}(2017)\citenamefont {Rafiee},
  \citenamefont {Nourmandipour},\ and\ \citenamefont
  {Mancini}}]{recenttwoqubit1}%
  \BibitemOpen
  \bibfield  {author} {\bibinfo {author} {\bibfnamefont {M.}~\bibnamefont
  {Rafiee}}, \bibinfo {author} {\bibfnamefont {A.}~\bibnamefont
  {Nourmandipour}}, \ and\ \bibinfo {author} {\bibfnamefont {S.}~\bibnamefont
  {Mancini}},\ }\href {\doibase 10.1103/PhysRevA.96.012340} {\bibfield
  {journal} {\bibinfo  {journal} {Phys. Rev. A}\ }\textbf {\bibinfo {volume}
  {96}},\ \bibinfo {pages} {012340} (\bibinfo {year} {2017})}\BibitemShut
  {NoStop}%
\bibitem [{\citenamefont {Bartkiewicz}\ \emph {et~al.}(2017)\citenamefont
  {Bartkiewicz}, \citenamefont {Chimczak},\ and\ \citenamefont
  {Lemr}}]{recenttwoqubit2}%
  \BibitemOpen
  \bibfield  {author} {\bibinfo {author} {\bibfnamefont {K.}~\bibnamefont
  {Bartkiewicz}}, \bibinfo {author} {\bibfnamefont {G.}~\bibnamefont
  {Chimczak}}, \ and\ \bibinfo {author} {\bibfnamefont {K.}~\bibnamefont
  {Lemr}},\ }\href {\doibase 10.1103/PhysRevA.95.022331} {\bibfield  {journal}
  {\bibinfo  {journal} {Phys. Rev. A}\ }\textbf {\bibinfo {volume} {95}},\
  \bibinfo {pages} {022331} (\bibinfo {year} {2017})}\BibitemShut {NoStop}%
\bibitem [{\citenamefont {Adhikari}(2018)}]{adhikari18}%
  \BibitemOpen
  \bibfield  {author} {\bibinfo {author} {\bibfnamefont {S.}~\bibnamefont
  {Adhikari}},\ }\href {\doibase 10.1103/PhysRevA.97.042344} {\bibfield
  {journal} {\bibinfo  {journal} {Phys. Rev. A}\ }\textbf {\bibinfo {volume}
  {97}},\ \bibinfo {pages} {042344} (\bibinfo {year} {2018})}\BibitemShut
  {NoStop}%
\bibitem [{\citenamefont {Yu}\ \emph {et~al.}(2008)\citenamefont {Yu},
  \citenamefont {Li},\ and\ \citenamefont {Song}}]{ent_wit_rank2}%
  \BibitemOpen
  \bibfield  {author} {\bibinfo {author} {\bibfnamefont {C.-s.}\ \bibnamefont
  {Yu}}, \bibinfo {author} {\bibfnamefont {C.}~\bibnamefont {Li}}, \ and\
  \bibinfo {author} {\bibfnamefont {H.-s.}\ \bibnamefont {Song}},\ }\href
  {\doibase 10.1103/PhysRevA.77.012305} {\bibfield  {journal} {\bibinfo
  {journal} {Phys. Rev. A}\ }\textbf {\bibinfo {volume} {77}},\ \bibinfo
  {pages} {012305} (\bibinfo {year} {2008})}\BibitemShut {NoStop}%
\bibitem [{\citenamefont {Augusiak}\ \emph {et~al.}(2008)\citenamefont
  {Augusiak}, \citenamefont {Demianowicz},\ and\ \citenamefont
  {Horodecki}}]{ADH_08}%
  \BibitemOpen
  \bibfield  {author} {\bibinfo {author} {\bibfnamefont {R.}~\bibnamefont
  {Augusiak}}, \bibinfo {author} {\bibfnamefont {M.}~\bibnamefont
  {Demianowicz}}, \ and\ \bibinfo {author} {\bibfnamefont {P.}~\bibnamefont
  {Horodecki}},\ }\href {\doibase 10.1103/PhysRevA.77.030301} {\bibfield
  {journal} {\bibinfo  {journal} {Phys. Rev. A}\ }\textbf {\bibinfo {volume}
  {77}},\ \bibinfo {pages} {030301(R)} (\bibinfo {year} {2008})}\BibitemShut
  {NoStop}%
\bibitem [{\citenamefont {Girolami}\ and\ \citenamefont
  {Adesso}(2012)}]{girolami_2012}%
  \BibitemOpen
  \bibfield  {author} {\bibinfo {author} {\bibfnamefont {D.}~\bibnamefont
  {Girolami}}\ and\ \bibinfo {author} {\bibfnamefont {G.}~\bibnamefont
  {Adesso}},\ }\href {\doibase 10.1103/PhysRevLett.108.150403} {\bibfield
  {journal} {\bibinfo  {journal} {Phys. Rev. Lett.}\ }\textbf {\bibinfo
  {volume} {108}},\ \bibinfo {pages} {150403} (\bibinfo {year}
  {2012})}\BibitemShut {NoStop}%
\bibitem [{\citenamefont {Zhang}\ \emph {et~al.}(2008)\citenamefont {Zhang},
  \citenamefont {Gong}, \citenamefont {Zhang},\ and\ \citenamefont
  {Guo}}]{twocopy_concurrence_mixed3}%
  \BibitemOpen
  \bibfield  {author} {\bibinfo {author} {\bibfnamefont {C.-J.}\ \bibnamefont
  {Zhang}}, \bibinfo {author} {\bibfnamefont {Y.-X.}\ \bibnamefont {Gong}},
  \bibinfo {author} {\bibfnamefont {Y.-S.}\ \bibnamefont {Zhang}}, \ and\
  \bibinfo {author} {\bibfnamefont {G.-C.}\ \bibnamefont {Guo}},\ }\href
  {\doibase 10.1103/PhysRevA.78.042308} {\bibfield  {journal} {\bibinfo
  {journal} {Phys. Rev. A}\ }\textbf {\bibinfo {volume} {78}},\ \bibinfo
  {pages} {042308} (\bibinfo {year} {2008})}\BibitemShut {NoStop}%
\bibitem [{\citenamefont {Lu}\ \emph {et~al.}(2016)\citenamefont {Lu},
  \citenamefont {Xin}, \citenamefont {Yu}, \citenamefont {Ji}, \citenamefont
  {Chen}, \citenamefont {Long}, \citenamefont {Baugh}, \citenamefont {Peng},
  \citenamefont {Zeng},\ and\ \citenamefont {Laflamme}}]{dawei}%
  \BibitemOpen
  \bibfield  {author} {\bibinfo {author} {\bibfnamefont {D.}~\bibnamefont
  {Lu}}, \bibinfo {author} {\bibfnamefont {T.}~\bibnamefont {Xin}}, \bibinfo
  {author} {\bibfnamefont {N.}~\bibnamefont {Yu}}, \bibinfo {author}
  {\bibfnamefont {Z.}~\bibnamefont {Ji}}, \bibinfo {author} {\bibfnamefont
  {J.}~\bibnamefont {Chen}}, \bibinfo {author} {\bibfnamefont {G.}~\bibnamefont
  {Long}}, \bibinfo {author} {\bibfnamefont {J.}~\bibnamefont {Baugh}},
  \bibinfo {author} {\bibfnamefont {X.}~\bibnamefont {Peng}}, \bibinfo {author}
  {\bibfnamefont {B.}~\bibnamefont {Zeng}}, \ and\ \bibinfo {author}
  {\bibfnamefont {R.}~\bibnamefont {Laflamme}},\ }\href {\doibase
  10.1103/PhysRevLett.116.230501} {\bibfield  {journal} {\bibinfo  {journal}
  {Phys. Rev. Lett.}\ }\textbf {\bibinfo {volume} {116}},\ \bibinfo {pages}
  {230501} (\bibinfo {year} {2016})}\BibitemShut {NoStop}%
\bibitem [{\citenamefont {Carmeli}\ \emph {et~al.}(2016)\citenamefont
  {Carmeli}, \citenamefont {Heinosaari}, \citenamefont {Karlsson},
  \citenamefont {Schultz},\ and\ \citenamefont {Toigo}}]{heinosaari_2012}%
  \BibitemOpen
  \bibfield  {author} {\bibinfo {author} {\bibfnamefont {C.}~\bibnamefont
  {Carmeli}}, \bibinfo {author} {\bibfnamefont {T.}~\bibnamefont {Heinosaari}},
  \bibinfo {author} {\bibfnamefont {A.}~\bibnamefont {Karlsson}}, \bibinfo
  {author} {\bibfnamefont {J.}~\bibnamefont {Schultz}}, \ and\ \bibinfo
  {author} {\bibfnamefont {A.}~\bibnamefont {Toigo}},\ }\href {\doibase
  10.1103/PhysRevLett.116.230403} {\bibfield  {journal} {\bibinfo  {journal}
  {Phys. Rev. Lett.}\ }\textbf {\bibinfo {volume} {116}},\ \bibinfo {pages}
  {230403} (\bibinfo {year} {2016})}\BibitemShut {NoStop}%
\bibitem [{\citenamefont {Mallick}\ and\ \citenamefont {Ghosh}(2017)}]{MG_17}%
  \BibitemOpen
  \bibfield  {author} {\bibinfo {author} {\bibfnamefont {A.}~\bibnamefont
  {Mallick}}\ and\ \bibinfo {author} {\bibfnamefont {S.}~\bibnamefont
  {Ghosh}},\ }\href {\doibase 10.1103/PhysRevA.96.052323} {\bibfield  {journal}
  {\bibinfo  {journal} {Phys. Rev. A}\ }\textbf {\bibinfo {volume} {96}},\
  \bibinfo {pages} {052323} (\bibinfo {year} {2017})}\BibitemShut {NoStop}%
\bibitem [{\citenamefont {Aharonov}\ \emph {et~al.}(1988)\citenamefont
  {Aharonov}, \citenamefont {Albert},\ and\ \citenamefont
  {Vaidman}}]{Vaidman_weak}%
  \BibitemOpen
  \bibfield  {author} {\bibinfo {author} {\bibfnamefont {Y.}~\bibnamefont
  {Aharonov}}, \bibinfo {author} {\bibfnamefont {D.~Z.}\ \bibnamefont
  {Albert}}, \ and\ \bibinfo {author} {\bibfnamefont {L.}~\bibnamefont
  {Vaidman}},\ }\href {\doibase 10.1103/PhysRevLett.60.1351} {\bibfield
  {journal} {\bibinfo  {journal} {Phys. Rev. Lett.}\ }\textbf {\bibinfo
  {volume} {60}},\ \bibinfo {pages} {1351} (\bibinfo {year}
  {1988})}\BibitemShut {NoStop}%
\bibitem [{\citenamefont {Hosten}\ and\ \citenamefont {Kwiat}(2008)}]{HK_08}%
  \BibitemOpen
  \bibfield  {author} {\bibinfo {author} {\bibfnamefont {O.}~\bibnamefont
  {Hosten}}\ and\ \bibinfo {author} {\bibfnamefont {P.}~\bibnamefont {Kwiat}},\
  }\href@noop {} {\bibfield  {journal} {\bibinfo  {journal} {Science}\ }\textbf
  {\bibinfo {volume} {319}},\ \bibinfo {pages} {787} (\bibinfo {year}
  {2008})}\BibitemShut {NoStop}%
\bibitem [{\citenamefont {Kocsis}\ \emph {et~al.}(2011)\citenamefont {Kocsis},
  \citenamefont {Braverman}, \citenamefont {Ravets}, \citenamefont {Stevens},
  \citenamefont {Mirin}, \citenamefont {Shalm},\ and\ \citenamefont
  {Steinberg}}]{KBRSM_11}%
  \BibitemOpen
  \bibfield  {author} {\bibinfo {author} {\bibfnamefont {S.}~\bibnamefont
  {Kocsis}}, \bibinfo {author} {\bibfnamefont {B.}~\bibnamefont {Braverman}},
  \bibinfo {author} {\bibfnamefont {S.}~\bibnamefont {Ravets}}, \bibinfo
  {author} {\bibfnamefont {M.~J.}\ \bibnamefont {Stevens}}, \bibinfo {author}
  {\bibfnamefont {R.~P.}\ \bibnamefont {Mirin}}, \bibinfo {author}
  {\bibfnamefont {L.~K.}\ \bibnamefont {Shalm}}, \ and\ \bibinfo {author}
  {\bibfnamefont {A.~M.}\ \bibnamefont {Steinberg}},\ }\href@noop {} {\bibfield
   {journal} {\bibinfo  {journal} {Science}\ }\textbf {\bibinfo {volume}
  {332}},\ \bibinfo {pages} {1170} (\bibinfo {year} {2011})}\BibitemShut
  {NoStop}%
\bibitem [{\citenamefont {Lundeen}\ \emph {et~al.}(2011)\citenamefont
  {Lundeen}, \citenamefont {Sutherland}, \citenamefont {Patel}, \citenamefont
  {Stewart},\ and\ \citenamefont {Bamber}}]{LSPSB_11}%
  \BibitemOpen
  \bibfield  {author} {\bibinfo {author} {\bibfnamefont {J.~S.}\ \bibnamefont
  {Lundeen}}, \bibinfo {author} {\bibfnamefont {B.}~\bibnamefont {Sutherland}},
  \bibinfo {author} {\bibfnamefont {A.}~\bibnamefont {Patel}}, \bibinfo
  {author} {\bibfnamefont {C.}~\bibnamefont {Stewart}}, \ and\ \bibinfo
  {author} {\bibfnamefont {C.}~\bibnamefont {Bamber}},\ }\href@noop {}
  {\bibfield  {journal} {\bibinfo  {journal} {Nature}\ }\textbf {\bibinfo
  {volume} {474}},\ \bibinfo {pages} {188} (\bibinfo {year}
  {2011})}\BibitemShut {NoStop}%
\bibitem [{\citenamefont {Str{\"u}bi}\ and\ \citenamefont
  {Bruder}(2013)}]{SB_13}%
  \BibitemOpen
  \bibfield  {author} {\bibinfo {author} {\bibfnamefont {G.}~\bibnamefont
  {Str{\"u}bi}}\ and\ \bibinfo {author} {\bibfnamefont {C.}~\bibnamefont
  {Bruder}},\ }\href@noop {} {\bibfield  {journal} {\bibinfo  {journal}
  {Physical review letters}\ }\textbf {\bibinfo {volume} {110}},\ \bibinfo
  {pages} {083605} (\bibinfo {year} {2013})}\BibitemShut {NoStop}%
\bibitem [{\citenamefont {Koashi}\ and\ \citenamefont {Ueda}(1999)}]{KU_99}%
  \BibitemOpen
  \bibfield  {author} {\bibinfo {author} {\bibfnamefont {M.}~\bibnamefont
  {Koashi}}\ and\ \bibinfo {author} {\bibfnamefont {M.}~\bibnamefont {Ueda}},\
  }\href@noop {} {\bibfield  {journal} {\bibinfo  {journal} {Physical review
  letters}\ }\textbf {\bibinfo {volume} {82}},\ \bibinfo {pages} {2598}
  (\bibinfo {year} {1999})}\BibitemShut {NoStop}%
\bibitem [{\citenamefont {Korotkov}\ and\ \citenamefont
  {Jordan}(2006)}]{KJ_06}%
  \BibitemOpen
  \bibfield  {author} {\bibinfo {author} {\bibfnamefont {A.~N.}\ \bibnamefont
  {Korotkov}}\ and\ \bibinfo {author} {\bibfnamefont {A.~N.}\ \bibnamefont
  {Jordan}},\ }\href@noop {} {\bibfield  {journal} {\bibinfo  {journal}
  {Physical review letters}\ }\textbf {\bibinfo {volume} {97}},\ \bibinfo
  {pages} {166805} (\bibinfo {year} {2006})}\BibitemShut {NoStop}%
\bibitem [{\citenamefont {Kim}\ \emph {et~al.}(2012)\citenamefont {Kim},
  \citenamefont {Lee}, \citenamefont {Kwon},\ and\ \citenamefont
  {Kim}}]{KLKK_12}%
  \BibitemOpen
  \bibfield  {author} {\bibinfo {author} {\bibfnamefont {Y.-S.}\ \bibnamefont
  {Kim}}, \bibinfo {author} {\bibfnamefont {J.-C.}\ \bibnamefont {Lee}},
  \bibinfo {author} {\bibfnamefont {O.}~\bibnamefont {Kwon}}, \ and\ \bibinfo
  {author} {\bibfnamefont {Y.-H.}\ \bibnamefont {Kim}},\ }\href@noop {}
  {\bibfield  {journal} {\bibinfo  {journal} {Nature Physics}\ }\textbf
  {\bibinfo {volume} {8}},\ \bibinfo {pages} {117} (\bibinfo {year}
  {2012})}\BibitemShut {NoStop}%
\bibitem [{\citenamefont {Man}\ \emph {et~al.}(2012)\citenamefont {Man},
  \citenamefont {Xia},\ and\ \citenamefont {An}}]{MXA_12}%
  \BibitemOpen
  \bibfield  {author} {\bibinfo {author} {\bibfnamefont {Z.-X.}\ \bibnamefont
  {Man}}, \bibinfo {author} {\bibfnamefont {Y.-J.}\ \bibnamefont {Xia}}, \ and\
  \bibinfo {author} {\bibfnamefont {N.~B.}\ \bibnamefont {An}},\ }\href@noop {}
  {\bibfield  {journal} {\bibinfo  {journal} {Physical Review A}\ }\textbf
  {\bibinfo {volume} {86}},\ \bibinfo {pages} {052322} (\bibinfo {year}
  {2012})}\BibitemShut {NoStop}%
\bibitem [{\citenamefont {Pramanik}\ and\ \citenamefont
  {Majumdar}(2013)}]{PM_13}%
  \BibitemOpen
  \bibfield  {author} {\bibinfo {author} {\bibfnamefont {T.}~\bibnamefont
  {Pramanik}}\ and\ \bibinfo {author} {\bibfnamefont {A.}~\bibnamefont
  {Majumdar}},\ }\href@noop {} {\bibfield  {journal} {\bibinfo  {journal}
  {Physics Letters A}\ }\textbf {\bibinfo {volume} {377}},\ \bibinfo {pages}
  {3209} (\bibinfo {year} {2013})}\BibitemShut {NoStop}%
\bibitem [{\citenamefont {Datta}\ \emph {et~al.}(2017)\citenamefont {Datta},
  \citenamefont {Goswami}, \citenamefont {Pramanik},\ and\ \citenamefont
  {Majumdar}}]{DGPM_17}%
  \BibitemOpen
  \bibfield  {author} {\bibinfo {author} {\bibfnamefont {S.}~\bibnamefont
  {Datta}}, \bibinfo {author} {\bibfnamefont {S.}~\bibnamefont {Goswami}},
  \bibinfo {author} {\bibfnamefont {T.}~\bibnamefont {Pramanik}}, \ and\
  \bibinfo {author} {\bibfnamefont {A.}~\bibnamefont {Majumdar}},\ }\href@noop
  {} {\bibfield  {journal} {\bibinfo  {journal} {Physics Letters A}\ }\textbf
  {\bibinfo {volume} {381}},\ \bibinfo {pages} {897} (\bibinfo {year}
  {2017})}\BibitemShut {NoStop}%
\bibitem [{\citenamefont {Sj{\"o}qvist}(2006)}]{w_value_geometric}%
  \BibitemOpen
  \bibfield  {author} {\bibinfo {author} {\bibfnamefont {E.}~\bibnamefont
  {Sj{\"o}qvist}},\ }\href@noop {} {\bibfield  {journal} {\bibinfo  {journal}
  {Physics Letters A}\ }\textbf {\bibinfo {volume} {359}},\ \bibinfo {pages}
  {187} (\bibinfo {year} {2006})}\BibitemShut {NoStop}%
\bibitem [{\citenamefont {Pati}\ \emph {et~al.}(2015)\citenamefont {Pati},
  \citenamefont {Singh},\ and\ \citenamefont {Sinha}}]{w_value-non-Herm}%
  \BibitemOpen
  \bibfield  {author} {\bibinfo {author} {\bibfnamefont {A.~K.}\ \bibnamefont
  {Pati}}, \bibinfo {author} {\bibfnamefont {U.}~\bibnamefont {Singh}}, \ and\
  \bibinfo {author} {\bibfnamefont {U.}~\bibnamefont {Sinha}},\ }\href
  {\doibase 10.1103/PhysRevA.92.052120} {\bibfield  {journal} {\bibinfo
  {journal} {Phys. Rev. A}\ }\textbf {\bibinfo {volume} {92}},\ \bibinfo
  {pages} {052120} (\bibinfo {year} {2015})}\BibitemShut {NoStop}%
\bibitem [{\citenamefont {Lundeen}\ and\ \citenamefont
  {Bamber}(2012)}]{w_value_state1}%
  \BibitemOpen
  \bibfield  {author} {\bibinfo {author} {\bibfnamefont {J.~S.}\ \bibnamefont
  {Lundeen}}\ and\ \bibinfo {author} {\bibfnamefont {C.}~\bibnamefont
  {Bamber}},\ }\href {\doibase 10.1103/PhysRevLett.108.070402} {\bibfield
  {journal} {\bibinfo  {journal} {Phys. Rev. Lett.}\ }\textbf {\bibinfo
  {volume} {108}},\ \bibinfo {pages} {070402} (\bibinfo {year}
  {2012})}\BibitemShut {NoStop}%
\bibitem [{\citenamefont {Haapasalo}\ \emph {et~al.}(2011)\citenamefont
  {Haapasalo}, \citenamefont {Lahti},\ and\ \citenamefont
  {Schultz}}]{w_value_state2}%
  \BibitemOpen
  \bibfield  {author} {\bibinfo {author} {\bibfnamefont {E.}~\bibnamefont
  {Haapasalo}}, \bibinfo {author} {\bibfnamefont {P.}~\bibnamefont {Lahti}}, \
  and\ \bibinfo {author} {\bibfnamefont {J.}~\bibnamefont {Schultz}},\ }\href
  {\doibase 10.1103/PhysRevA.84.052107} {\bibfield  {journal} {\bibinfo
  {journal} {Phys. Rev. A}\ }\textbf {\bibinfo {volume} {84}},\ \bibinfo
  {pages} {052107} (\bibinfo {year} {2011})}\BibitemShut {NoStop}%
\bibitem [{\citenamefont {Malik}\ \emph {et~al.}(2014)\citenamefont {Malik},
  \citenamefont {Mirhosseini}, \citenamefont {Lavery}, \citenamefont {Leach},
  \citenamefont {Padgett},\ and\ \citenamefont {Boyd}}]{w_value_state3}%
  \BibitemOpen
  \bibfield  {author} {\bibinfo {author} {\bibfnamefont {M.}~\bibnamefont
  {Malik}}, \bibinfo {author} {\bibfnamefont {M.}~\bibnamefont {Mirhosseini}},
  \bibinfo {author} {\bibfnamefont {M.~P.}\ \bibnamefont {Lavery}}, \bibinfo
  {author} {\bibfnamefont {J.}~\bibnamefont {Leach}}, \bibinfo {author}
  {\bibfnamefont {M.~J.}\ \bibnamefont {Padgett}}, \ and\ \bibinfo {author}
  {\bibfnamefont {R.~W.}\ \bibnamefont {Boyd}},\ }\href@noop {} {\bibfield
  {journal} {\bibinfo  {journal} {Nature communications}\ }\textbf {\bibinfo
  {volume} {5}},\ \bibinfo {pages} {3115} (\bibinfo {year} {2014})}\BibitemShut
  {NoStop}%
\bibitem [{\citenamefont {Mintert}\ \emph {et~al.}(2005)\citenamefont
  {Mintert}, \citenamefont {Ku\ifmmode~\acute{s}\else \'{s}\fi{}},\ and\
  \citenamefont {Buchleitner}}]{twocopy_concurrence_pure}%
  \BibitemOpen
  \bibfield  {author} {\bibinfo {author} {\bibfnamefont {F.}~\bibnamefont
  {Mintert}}, \bibinfo {author} {\bibfnamefont {M.}~\bibnamefont
  {Ku\ifmmode~\acute{s}\else \'{s}\fi{}}}, \ and\ \bibinfo {author}
  {\bibfnamefont {A.}~\bibnamefont {Buchleitner}},\ }\href {\doibase
  10.1103/PhysRevLett.95.260502} {\bibfield  {journal} {\bibinfo  {journal}
  {Phys. Rev. Lett.}\ }\textbf {\bibinfo {volume} {95}},\ \bibinfo {pages}
  {260502} (\bibinfo {year} {2005})}\BibitemShut {NoStop}%
\bibitem [{\citenamefont {Mintert}\ and\ \citenamefont
  {Buchleitner}(2007)}]{twocopy_concurrence_mixed1}%
  \BibitemOpen
  \bibfield  {author} {\bibinfo {author} {\bibfnamefont {F.}~\bibnamefont
  {Mintert}}\ and\ \bibinfo {author} {\bibfnamefont {A.}~\bibnamefont
  {Buchleitner}},\ }\href {\doibase 10.1103/PhysRevLett.98.140505} {\bibfield
  {journal} {\bibinfo  {journal} {Phys. Rev. Lett.}\ }\textbf {\bibinfo
  {volume} {98}},\ \bibinfo {pages} {140505} (\bibinfo {year}
  {2007})}\BibitemShut {NoStop}%
\bibitem [{\citenamefont {Schmid}\ \emph {et~al.}(2008)\citenamefont {Schmid},
  \citenamefont {Kiesel}, \citenamefont {Wieczorek}, \citenamefont
  {Weinfurter}, \citenamefont {Mintert},\ and\ \citenamefont
  {Buchleitner}}]{twocopy_concurrence_mixed2}%
  \BibitemOpen
  \bibfield  {author} {\bibinfo {author} {\bibfnamefont {C.}~\bibnamefont
  {Schmid}}, \bibinfo {author} {\bibfnamefont {N.}~\bibnamefont {Kiesel}},
  \bibinfo {author} {\bibfnamefont {W.}~\bibnamefont {Wieczorek}}, \bibinfo
  {author} {\bibfnamefont {H.}~\bibnamefont {Weinfurter}}, \bibinfo {author}
  {\bibfnamefont {F.}~\bibnamefont {Mintert}}, \ and\ \bibinfo {author}
  {\bibfnamefont {A.}~\bibnamefont {Buchleitner}},\ }\href {\doibase
  10.1103/PhysRevLett.101.260505} {\bibfield  {journal} {\bibinfo  {journal}
  {Phys. Rev. Lett.}\ }\textbf {\bibinfo {volume} {101}},\ \bibinfo {pages}
  {260505} (\bibinfo {year} {2008})}\BibitemShut {NoStop}%
\bibitem [{\citenamefont {Coffman}\ \emph {et~al.}(2000)\citenamefont
  {Coffman}, \citenamefont {Kundu},\ and\ \citenamefont {Wootters}}]{coffman}%
  \BibitemOpen
  \bibfield  {author} {\bibinfo {author} {\bibfnamefont {V.}~\bibnamefont
  {Coffman}}, \bibinfo {author} {\bibfnamefont {J.}~\bibnamefont {Kundu}}, \
  and\ \bibinfo {author} {\bibfnamefont {W.~K.}\ \bibnamefont {Wootters}},\
  }\href {\doibase 10.1103/PhysRevA.61.052306} {\bibfield  {journal} {\bibinfo
  {journal} {Phys. Rev. A}\ }\textbf {\bibinfo {volume} {61}},\ \bibinfo
  {pages} {052306} (\bibinfo {year} {2000})}\BibitemShut {NoStop}%
\bibitem [{\citenamefont {Wootters}(1998)}]{wootters_concurrence}%
  \BibitemOpen
  \bibfield  {author} {\bibinfo {author} {\bibfnamefont {W.~K.}\ \bibnamefont
  {Wootters}},\ }\href {\doibase 10.1103/PhysRevLett.80.2245} {\bibfield
  {journal} {\bibinfo  {journal} {Phys. Rev. Lett.}\ }\textbf {\bibinfo
  {volume} {80}},\ \bibinfo {pages} {2245} (\bibinfo {year}
  {1998})}\BibitemShut {NoStop}%
\bibitem [{\citenamefont {Sanpera}\ \emph {et~al.}(1998)\citenamefont
  {Sanpera}, \citenamefont {Tarrach},\ and\ \citenamefont
  {Vidal}}]{sanpera_two_qubit}%
  \BibitemOpen
  \bibfield  {author} {\bibinfo {author} {\bibfnamefont {A.}~\bibnamefont
  {Sanpera}}, \bibinfo {author} {\bibfnamefont {R.}~\bibnamefont {Tarrach}}, \
  and\ \bibinfo {author} {\bibfnamefont {G.}~\bibnamefont {Vidal}},\ }\href
  {\doibase 10.1103/PhysRevA.58.826} {\bibfield  {journal} {\bibinfo  {journal}
  {Phys. Rev. A}\ }\textbf {\bibinfo {volume} {58}},\ \bibinfo {pages} {826}
  (\bibinfo {year} {1998})}\BibitemShut {NoStop}%
\bibitem [{\citenamefont {Verstraete}\ \emph {et~al.}(2001)\citenamefont
  {Verstraete}, \citenamefont {Audenaert}, \citenamefont {Dehaene},\ and\
  \citenamefont {Moor}}]{Verstraete_two_qubit}%
  \BibitemOpen
  \bibfield  {author} {\bibinfo {author} {\bibfnamefont {F.}~\bibnamefont
  {Verstraete}}, \bibinfo {author} {\bibfnamefont {K.}~\bibnamefont
  {Audenaert}}, \bibinfo {author} {\bibfnamefont {J.}~\bibnamefont {Dehaene}},
  \ and\ \bibinfo {author} {\bibfnamefont {B.~D.}\ \bibnamefont {Moor}},\
  }\href {http://stacks.iop.org/0305-4470/34/i=47/a=329} {\bibfield  {journal}
  {\bibinfo  {journal} {Journal of Physics A: Mathematical and General}\
  }\textbf {\bibinfo {volume} {34}},\ \bibinfo {pages} {10327} (\bibinfo {year}
  {2001})}\BibitemShut {NoStop}%
\bibitem [{\citenamefont {Dressel}\ \emph {et~al.}(2014)\citenamefont
  {Dressel}, \citenamefont {Malik}, \citenamefont {Miatto}, \citenamefont
  {Jordan},\ and\ \citenamefont {Boyd}}]{weak_review}%
  \BibitemOpen
  \bibfield  {author} {\bibinfo {author} {\bibfnamefont {J.}~\bibnamefont
  {Dressel}}, \bibinfo {author} {\bibfnamefont {M.}~\bibnamefont {Malik}},
  \bibinfo {author} {\bibfnamefont {F.~M.}\ \bibnamefont {Miatto}}, \bibinfo
  {author} {\bibfnamefont {A.~N.}\ \bibnamefont {Jordan}}, \ and\ \bibinfo
  {author} {\bibfnamefont {R.~W.}\ \bibnamefont {Boyd}},\ }\href {\doibase
  10.1103/RevModPhys.86.307} {\bibfield  {journal} {\bibinfo  {journal} {Rev.
  Mod. Phys.}\ }\textbf {\bibinfo {volume} {86}},\ \bibinfo {pages} {307}
  (\bibinfo {year} {2014})}\BibitemShut {NoStop}%
\bibitem [{\citenamefont {Jozsa}(2007)}]{Jozsacomplex}%
  \BibitemOpen
  \bibfield  {author} {\bibinfo {author} {\bibfnamefont {R.}~\bibnamefont
  {Jozsa}},\ }\href {\doibase 10.1103/PhysRevA.76.044103} {\bibfield  {journal}
  {\bibinfo  {journal} {Phys. Rev. A}\ }\textbf {\bibinfo {volume} {76}},\
  \bibinfo {pages} {044103} (\bibinfo {year} {2007})}\BibitemShut {NoStop}%
\bibitem [{\citenamefont {Kobayashi}\ \emph {et~al.}(2014)\citenamefont
  {Kobayashi}, \citenamefont {Nonaka},\ and\ \citenamefont
  {Shikano}}]{w_value_detection_single}%
  \BibitemOpen
  \bibfield  {author} {\bibinfo {author} {\bibfnamefont {H.}~\bibnamefont
  {Kobayashi}}, \bibinfo {author} {\bibfnamefont {K.}~\bibnamefont {Nonaka}}, \
  and\ \bibinfo {author} {\bibfnamefont {Y.}~\bibnamefont {Shikano}},\ }\href
  {\doibase 10.1103/PhysRevA.89.053816} {\bibfield  {journal} {\bibinfo
  {journal} {Phys. Rev. A}\ }\textbf {\bibinfo {volume} {89}},\ \bibinfo
  {pages} {053816} (\bibinfo {year} {2014})}\BibitemShut {NoStop}%
\bibitem [{\citenamefont {Tukiainen}\ \emph {et~al.}(2017)\citenamefont
  {Tukiainen}, \citenamefont {Kobayashi},\ and\ \citenamefont
  {Shikano}}]{tukia_weak}%
  \BibitemOpen
  \bibfield  {author} {\bibinfo {author} {\bibfnamefont {M.}~\bibnamefont
  {Tukiainen}}, \bibinfo {author} {\bibfnamefont {H.}~\bibnamefont
  {Kobayashi}}, \ and\ \bibinfo {author} {\bibfnamefont {Y.}~\bibnamefont
  {Shikano}},\ }\href {\doibase 10.1103/PhysRevA.95.052301} {\bibfield
  {journal} {\bibinfo  {journal} {Phys. Rev. A}\ }\textbf {\bibinfo {volume}
  {95}},\ \bibinfo {pages} {052301} (\bibinfo {year} {2017})}\BibitemShut
  {NoStop}%
\bibitem [{\citenamefont {Grassl}\ \emph {et~al.}(1998)\citenamefont {Grassl},
  \citenamefont {R\"otteler},\ and\ \citenamefont {Beth}}]{Grasslpolynomial}%
  \BibitemOpen
  \bibfield  {author} {\bibinfo {author} {\bibfnamefont {M.}~\bibnamefont
  {Grassl}}, \bibinfo {author} {\bibfnamefont {M.}~\bibnamefont {R\"otteler}},
  \ and\ \bibinfo {author} {\bibfnamefont {T.}~\bibnamefont {Beth}},\ }\href
  {\doibase 10.1103/PhysRevA.58.1833} {\bibfield  {journal} {\bibinfo
  {journal} {Phys. Rev. A}\ }\textbf {\bibinfo {volume} {58}},\ \bibinfo
  {pages} {1833} (\bibinfo {year} {1998})}\BibitemShut {NoStop}%
\bibitem [{\citenamefont {\ifmmode~\dot{Z}\else \.{Z}\fi{}yczkowski}\ \emph
  {et~al.}(1998)\citenamefont {\ifmmode~\dot{Z}\else \.{Z}\fi{}yczkowski},
  \citenamefont {Horodecki}, \citenamefont {Sanpera},\ and\ \citenamefont
  {Lewenstein}}]{negativity2}%
  \BibitemOpen
  \bibfield  {author} {\bibinfo {author} {\bibfnamefont {K.}~\bibnamefont
  {\ifmmode~\dot{Z}\else \.{Z}\fi{}yczkowski}}, \bibinfo {author}
  {\bibfnamefont {P.}~\bibnamefont {Horodecki}}, \bibinfo {author}
  {\bibfnamefont {A.}~\bibnamefont {Sanpera}}, \ and\ \bibinfo {author}
  {\bibfnamefont {M.}~\bibnamefont {Lewenstein}},\ }\href {\doibase
  10.1103/PhysRevA.58.883} {\bibfield  {journal} {\bibinfo  {journal} {Phys.
  Rev. A}\ }\textbf {\bibinfo {volume} {58}},\ \bibinfo {pages} {883} (\bibinfo
  {year} {1998})}\BibitemShut {NoStop}%
\bibitem [{cro()}]{crossprod}%
  \BibitemOpen
  \href@noop {} {}\bibinfo {note} {Here for any set $\{a,b\}$, we define the
  cartesian product as
  $\{a,b\}\times\{a,b\}\equiv\{a^2,~ab,~ba,~b^2\}$.}\BibitemShut {Stop}%
\bibitem [{\citenamefont {Bruschi}\ \emph {et~al.}(2015)\citenamefont
  {Bruschi}, \citenamefont {Perarnau-Llobet}, \citenamefont {Friis},
  \citenamefont {Hovhannisyan},\ and\ \citenamefont
  {Huber}}]{energycorrelation}%
  \BibitemOpen
  \bibfield  {author} {\bibinfo {author} {\bibfnamefont {D.~E.}\ \bibnamefont
  {Bruschi}}, \bibinfo {author} {\bibfnamefont {M.}~\bibnamefont
  {Perarnau-Llobet}}, \bibinfo {author} {\bibfnamefont {N.}~\bibnamefont
  {Friis}}, \bibinfo {author} {\bibfnamefont {K.~V.}\ \bibnamefont
  {Hovhannisyan}}, \ and\ \bibinfo {author} {\bibfnamefont {M.}~\bibnamefont
  {Huber}},\ }\href {\doibase 10.1103/PhysRevE.91.032118} {\bibfield  {journal}
  {\bibinfo  {journal} {Phys. Rev. E}\ }\textbf {\bibinfo {volume} {91}},\
  \bibinfo {pages} {032118} (\bibinfo {year} {2015})}\BibitemShut {NoStop}%
\bibitem [{\citenamefont {Abdelkhalek}\ \emph {et~al.}(2016)\citenamefont
  {Abdelkhalek}, \citenamefont {Nakata},\ and\ \citenamefont
  {Reeb}}]{energymeasurement1}%
  \BibitemOpen
  \bibfield  {author} {\bibinfo {author} {\bibfnamefont {K.}~\bibnamefont
  {Abdelkhalek}}, \bibinfo {author} {\bibfnamefont {Y.}~\bibnamefont {Nakata}},
  \ and\ \bibinfo {author} {\bibfnamefont {D.}~\bibnamefont {Reeb}},\
  }\href@noop {} {\bibfield  {journal} {\bibinfo  {journal} {arXiv preprint
  arXiv:1609.06981}\ } (\bibinfo {year} {2016})}\BibitemShut {NoStop}%
\bibitem [{\citenamefont {Navascu\'es}\ and\ \citenamefont
  {Popescu}(2014)}]{energymeasurement2}%
  \BibitemOpen
  \bibfield  {author} {\bibinfo {author} {\bibfnamefont {M.}~\bibnamefont
  {Navascu\'es}}\ and\ \bibinfo {author} {\bibfnamefont {S.}~\bibnamefont
  {Popescu}},\ }\href {\doibase 10.1103/PhysRevLett.112.140502} {\bibfield
  {journal} {\bibinfo  {journal} {Phys. Rev. Lett.}\ }\textbf {\bibinfo
  {volume} {112}},\ \bibinfo {pages} {140502} (\bibinfo {year}
  {2014})}\BibitemShut {NoStop}%
\end{thebibliography}%
\end{document}